\documentclass{ws-rv961x669}
\usepackage{subfigure}   
\usepackage{ws-rv-thm}   
\usepackage{ws-rv-van}   
\usepackage{ws-index}   
\makeindex

\usepackage{graphicx}
\usepackage{textcomp}
\usepackage{slashed}
\usepackage{amsmath}

\newcommand{\lsim}   {\mathrel{\mathop{\kern 0pt \rlap
  {\raise.2ex\hbox{$<$}}}
  \lower.9ex\hbox{\kern-.190em $\sim$}}}
\newcommand{\gsim}   {\mathrel{\mathop{\kern 0pt \rlap
  {\raise.2ex\hbox{$>$}}}
  \lower.9ex\hbox{\kern-.190em $\sim$}}}
\usepackage[
 colorlinks=true,    
 linkcolor=blue,     
 citecolor=blue,     
 filecolor=blue,  
 urlcolor=blue,      
 final=true
]{hyperref}

\usepackage[force]{feynmp-auto} 
\graphicspath{{figures-1/}} 

\newcommand\dslash{\,/\!\!\!\!D}
\newcommand\pslash{\,/\!\!\!p}
\newcommand\psl{\,/\!\!\!\partial}
\newcommand\bra[1]{\langle #1 |}
\newcommand\ket[1]{| #1 \rangle}
\newcommand\braket[2]{\langle #1 | #2 \rangle}

\def\theta{\vartheta}
\begin{document}



\title{Neutrino Physics and Astrophysics}

\chapter[Neutrino Properties]{Neutrino Properties and Interactions}

\author[Pedro A. N. Machado]{Pedro A. N. Machado\footnote{pmachado@fnal.gov}}
\address{Fermi National Accelerator Laboratory\\ Batavia, IL 60510-0500, U.S.A.}

\body

\abstract{In this chapter we present basic concepts of neutrino physics. We start with a brief introduction to the standard model electroweak sector, followed by calculations of some relevant neutrino interaction cross sections. We obtain the oscillation formalism from the Dirac equation, alongside with a self-contained derivation of matter effects from the standard model electroweak lagrangian. Some key features of neutrino oscillation phenomenology are discussed as well. We then review the most precise oscillation measurements and finalize with some broad discussion of the current open questions in neutrino physics.\footnote{To be published in ”Neutrino Physics and Astrophysics”, edited by F. W.
Stecker, in Encyclopedia of Cosmology II, edited by G. G. Fazio, World Scientific
Publishing Company, Singapore, 2022.}
}

\tableofcontents

\newpage

\section{Introduction}

The formulation of the standard model has been an intense, rich, and long-term endeavor of particle physics, receiving important theoretical and experimental contributions for over 60 years.
Nonetheless, the evidence and the proposal of the existence of the neutrino predates the standard model.
In this introduction we will briefly describe some outstanding aspects and open questions in the neutrino sector.
The attentive reader will notice that these aspects are related to each other in one way or another.
We hope this will motivate the reader to go through the exciting physics of neutrinos and their role in cosmology, particle and astroparticle physics.

The first evidence of the existence of neutrinos has its roots in early observations of beta decays by Sir James Chadwick in 1915~\cite{Chadwick:1914zz}, for which the observed continuous electron energy spectrum completely disagreed with a monochromatic electron energy as expected in two body decays.
The puzzle of beta decays culminated in the famous letter by Wolfgang Pauli~\cite{Pauli:1930pc} proposing the existence of an invisible particle which would turn out to be the neutrino.
Later, Fermi would write down a Hamiltonian describing the beta decay transition~\cite{Fermi:1934hr}, a major landmark in the early stages of the formulation of the standard model and in the understanding of neutrino interactions.

The neutrino-nuclei interaction cross section predicted by Fermi's four-fermion effective theory was extremely small.
As neutrinos have no electric charge neither color, their only known interactions happen via the weak gauge bosons.
Today, we know that the smallness of Fermi's constant, the dimensionful coupling in Fermi's theory, is due to a suppression coming from the large mass of the $W$ boson of about 80~GeV.
Being endowed  with only weak interactions is what makes neutrinos special but also the main obstacle to study these particles.
To give the reader a more quantitative idea, the inverse beta decay reaction $\bar\nu_e p^+\to e^+n$, used by the pioneering experiment of Cowan and Reines~\cite{Cowan:1992xc} which detected neutrinos (produced in nuclear reactors) for the first time, has a cross section of $\sigma_{\rm IBD}\sim10^{-43}(E_\nu/{\rm MeV})^2$~cm$^2$, where $E_\nu$ is the neutrino energy. 
This means that the mean free path of a 3~MeV electron antineutrino neutrino in a hypothetical background with a density of Avogadro's number of protons per cm$^3$ is $\lambda\sim10^{13}$~km, that is, 100,000 times the Earth-Sun distance!

In fact, neutrinos are so challenging to study that some standard interaction channels between neutrino and ordinary matter, despite having been discussed theoretically many years ago, have only been recently discovered. 
Two of those channels stand out. 
Low energy neutrinos may scatter off nuclei coherently, being sensitive to the overall weak charge of nuclei, as opposed to each individual nucleon.
Although the cross section is enhanced by the square of the weak charge of the nucleus, and thus typically referred to as a relatively large neutrino cross section, its experimental signature is very challenging.
To have coherence, the energy transferred to the nucleus cannot be much above the MeV scale.
This typically translates into keV nuclear recoils.
Disentangling such small deposits of energy from backgrounds induced by cosmic rays or neutrons is a difficult task.
Indeed, although this process has first been discussed in 1973~\cite{Freedman:1973yd}, it took over 40 years to be observed by the COHERENT experiment~\cite{Akimov:2017ade}.

In the high energy side, the resonant $W$ scattering, also known as the ``Glashow resonance'', was proposed in the 1960's~\cite{Glashow:1960zz}.
In this process, an ultra high energy electron antineutrino with energy around 6.3~PeV scatters off an electron producing a real $W$ boson which then decays to leptons or hadrons, namely $\bar\nu_e e^-\to W^-\to \text{anything}$.
Due to the resonant character of this process, this cross section is enhanced relatively to its non-resonant counterpart as in $\nu_e e^-\to\nu_e e^-$ by a factor of order $M_W^2/\Gamma_W^2\sim1600$, where $M_W$ and $\Gamma_W$ are the mass and width of the $W$ boson.
Nevertheless, such high-energy neutrinos can only have astrophysical origin, requiring gigantic neutrino observatories to be detected, as their flux is fairly low.
The IceCube Neutrino Observatory has detected an excess of events with an energy consistent with the Glashow resonance~\cite{IceCube:2021rpz}. However, larger sample sizes are warranted for a definite claim of the observation of a Glashow process interaction.
Neutrino interactions will be discussed in more detail in Sec.~\ref{sec:neutrino-interactions}.

An important feature of having only weak interactions is that neutrinos can freely travel through matter in almost all environments, with the exception of some extreme cases such as in supernovae.
From the tiny interaction rate of neutrinos with normal matter, plus the facts that they can be produced in weak decays and that they are the lightest fermions of the standard model follows their ubiquity in the universe and their importance in cosmology and astrophysics.
For instance, the decoupling of neutrinos from the plasma in the early universe plays a crucial role in the abundance of light elements sourced during big bang nucleosynthesis.
In supernovae, neutrinos are the first particle known to escape the dense environment and thus are critical in the dynamics of supernova explosions.
Last, the dominant cooling mechanism in stars, including our Sun, is by far the emission (and escape) of neutrinos. 

Regarding our own star, neutrinos emitted in the core of the Sun were observed in 1964 by Ray Davis~\cite{Davis:1964hf}, proving that neutrinos indeed provide the main mechanism for cooling down stars.
Nevertheless, the flux of solar neutrinos observed experimentally exhibited a large deficit with respect to theoretical expectations~\cite{Bahcall:1964gx}.
The solution to the so-called \emph{solar neutrino problem} revealed what is perhaps most distinctive feature of neutrinos: the phenomenon of neutrino oscillations.

In short, the oscillation of neutrinos from one flavor to another happens because weak interactions produce ``flavor eigenstates,'' which do not have a well defined mass. 
Instead, flavor states are quantum superpositions of ``mass states,'' which are eigenstates of the free Hamiltonian and therefore have well defined masses.
Propagation changes relative phases among mass states and consequently their projection into flavor states, leading to oscillations.
This effect has also been observed in the quark sector, though the amplitude of oscillations there,  parametrized by mixing angles, is found to be relatively small.
Neutrino oscillations, on the other hand, are governed by very large mixings, one of which may possibly be maximal.
The understanding of neutrino oscillations is among the most active fields of particle physics in the last 30 years.
We will discuss the theoretical framework of oscillations and related experimental measurements in Secs.~\ref{sec:oscillations} and \ref{sec:measurements}.
A crucial aspect of neutrino oscillations is that it can only occur if neutrinos have different masses, so that the neutrino wave packets travel at slightly different speeds and a relative phase develops. 
Due to that, the observation of oscillations is evidence, and the only one so far, that neutrinos have a nonzero masses~\cite{Kajita:2016cak,McDonald:2016ixn}. 

\begin{figure}[t]
\centering
\includegraphics[width=0.9\textwidth]{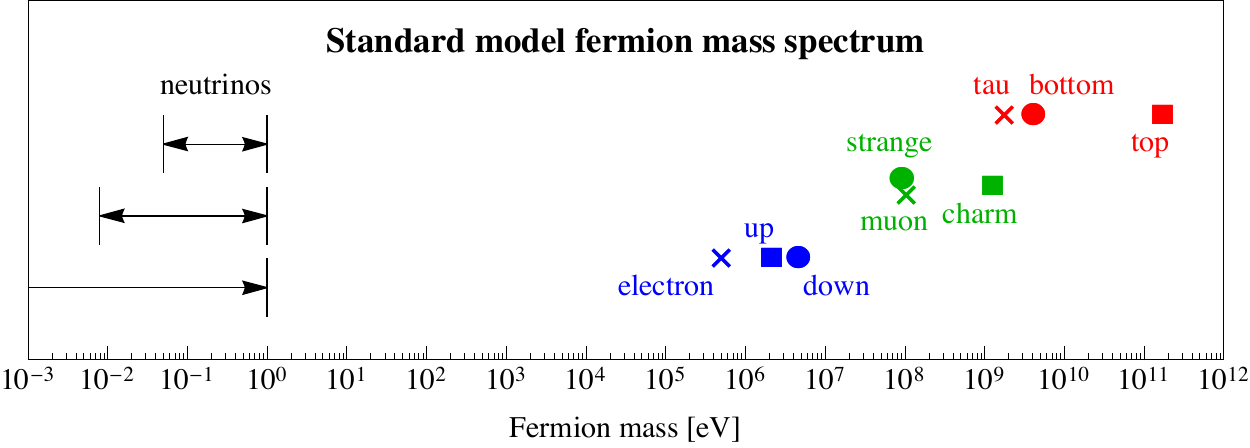}
\caption{\label{fig:fermion_spectrum}Mass spectrum of standard model fermions. Charged leptons, up-type quarks and  down-type quarks are indicated by ``$\times$,'' ``{\large$\bullet$},'' and ``$\blacksquare$,'' while first, second and third generations are depicted by blue, green and red, respectively.
 A rough estimate of the allowed mass region for the three neutrinos is indicated in red. Figure adapted from Ref.~\cite{deGouvea:2004gd}.}
\end{figure}

Neutrino masses are one of the outstanding problems of the standard model mainly due to  three reasons.
First, laboratory and cosmological constraints on neutrino masses indicate that they are not larger than about 1~eV. 
Compared to the top quark, this represents a difference of at least 11 orders of magnitude in fermion masses, see Fig.~\ref{fig:fermion_spectrum}.
This large difference makes one wonder if the same mechanism generates masses for all fermions, and if there is some rationale behind the fermion mass spectrum.
The lack of understanding of the pattern standard model fermion masses and mixings is typically referred to as the \emph{flavor puzzle}.

Second, if one just tries to repeat the Higgs mechanism (which will be discussed shortly) for neutrinos, a right-handed neutrino should be present in the particle spectrum. 
This particle has not been observed, and thus it stands to reason that its existence would be physics beyond the standard model.
The right-handed neutrino would carry no gauge quantum number whatsoever, and thus would not interact via strong, electromagnetic nor weak forces.
Due to that, this fermion  could have its own mass term, what does not happen to any other fermion in the standard model.
This could drastically affect the nature of neutrinos, making them Majorana particles, that is, the neutrino could be its own antiparticle. 
As a consequence, lepton number would be violated by neutrino masses, which could possibly be linked to the matter-antimatter asymmetry of the universe in so-called \emph{leptogenesis} scenarios~\cite{Fukugita:1986hr}.

Finally, as the mechanism of neutrino mass generation necessarily requires beyond standard model particles, there is no univocal way of realizing it.
The simplest models, the \emph{seesaw scenarios}, can be implemented in three different ways, while more ambitious theoretical frameworks such as grand unified theories may yield numerous possibilities for the neutrino mass generation.
We will discuss the puzzle of neutrino masses and simple seesaw models in Secs.~\ref{sec:standard-model} and \ref{sec:neutrino-mass-mechanism}, while excellent books can be found for grand unified theories and other models of neutrino mass generation (see, e.g. Ref.~\cite{Mohapatra:1998rq}).

The problem of neutrino masses is deeply connected to the chirality of weak interactions.
Although a historical description of the standard model is not the goal of this chapter, understanding a few key discoveries that led to the formulation of the standard model may help the reader to comprehend, or at least accept, the structure of the standard model.
One of those, is the discovery of parity violation in weak interactions via the observation of polarized $^{60}$Co decays in the Wu experiment~\cite{Wu:1957my} and via the decay chain of pions in the Garwin-Lederman-Weinrich experiment~\cite{Garwin:1957hc}.
In both experiments, the violation of parity is imprinted in the helicity and angular distributions of daughter particles in the relevant weak decay processes.

The violation of parity is a consequence of the fact that the weak force, as far as we know, only acts on left-handed fields.
Due to gauge invariance, the most striking consequence of such fact is the absence of ``bare'' mass term for fermions in the standard model. More precisely, terms like $m\bar\psi_L\psi_R$, where $\psi$ is a fermion and the subscripts ``$L$'' and ``$R$'' denote left- and right-handed chirality, are forbidden by weak interactions. 
The observation of nonzero fermion masses requires the presence of some mechanism to circumvent gauge invariance.

The mechanism of spontaneous symmetry breaking of the standard model, proposed concomitantly  by Higgs, Brout-Englert and Guralnik-Hagen-Kibble~\cite{Higgs:1964pj,Englert:1964et,Guralnik:1964eu}, most commonly known as the ``Higgs mechanism,'' is a way of generating nonzero fermion and weak gauge boson masses without violating the gauge symmetry.
In a nutshell, a scalar boson (the Higgs boson) is added to the standard model Lagrangian with a potential that has a minimum away from zero.
Although the gauge symmetry is still conserved, the Higgs boson acquires a vacuum expectation value that breaks the symmetry spontaneously. 
This is similar to ferromagnetism, in which the alignment of electron spins is energetically favorable (below a certain critical temperature), but the \emph{direction} of this alignment is random, as the laws describing ferromagnetism are invariant under spatial rotations.
Nevertheless, as a random direction is chosen by the alignment, rotation symmetry is spontaneously broken.
The discovery of a particle consistent with the Higgs boson by the CMS and ATLAS experiments at the Large Hadron Collider in 2012~\cite{ATLAS:2012yve,CMS:2012qbp} has essentially established that the Higgs mechanism is indeed what generates masses for standard model charged fermions, or at least those of the third family.

This chapter is organized as follows.
In Sec.~\ref{sec:standard-model} we will discuss the electroweak sector of the standard model, showing how the Higgs mechanism generates masses for weak bosons and fermions, and laying down the electroweak interactions of fermions.
Neutrino interactions in several energy regimes will be discussed in Sec.~\ref{sec:neutrino-interactions}.
Theoretical aspects of neutrino oscillations will be covered in Sec.~\ref{sec:oscillations} while experimental determinations of mass splittings and mixings will be found in Sec.~\ref{sec:measurements}.
Finally, in Sec.~\ref{sec:questions} we discuss some of the open questions in neutrino physics, including the neutrino mass mechanism, the Majorana versus Dirac nature of neutrinos, and experimental anomalies that remain unexplained.

\vspace{0.3cm}
\textbf{Conventions:}
We adopt the metric $g_{\mu\nu}={\rm diag}(+1\,\,-1\,\,-1\,\,-1)$, such that contracting the four momentum $p^\mu=(E,\,\boldsymbol{p})$ of a particle with itself yields $p^\mu p_\mu=E^2-\boldsymbol{p}^2=m^2$, where $m$ denotes the particle mass. Unless otherwise noted, we will use natural units throughout this chapter, $\hbar=c=1$.


\section{The standard model}\label{sec:standard-model}

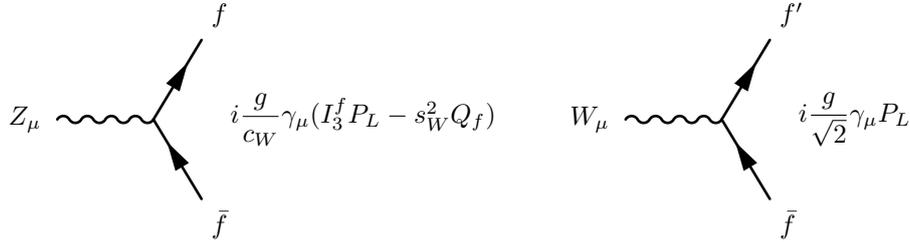
\begin{figure}[t]
\hspace{0.7cm}
\begin{fmffile}{diagram_Zff}
\begin{fmfgraph*}(60, 60)
	\fmfleft{i1}
	\fmfright{o1,o2}
	\fmflabel{$Z_\mu$}{i1}
	\fmflabel{$\bar f$}{o1}
	\fmflabel{$f$}{o2}
	\fmflabel{$\hspace{0.8cm}i\dfrac{g}{c_W}\gamma_\mu(I_3^fP_L-s^2_W Q_f)$}{v1}
	\fmf{fermion}{o1,v1,o2}
	\fmf{photon}{i1,v1}
\end{fmfgraph*}
\end{fmffile}\hspace{4.9cm} 
\begin{fmffile}{diagram_Wff}
\begin{fmfgraph*}(60, 60)
	\fmfleft{i1}
	\fmfright{o1,o2}
	\fmflabel{$W_\mu$}{i1}
	\fmflabel{$\bar f$}{o1}
	\fmflabel{$f'$}{o2}
	\fmflabel{$\hspace{0.8cm}i\dfrac{g}{\sqrt2}\gamma_\mu P_L$}{v1}
	\fmf{fermion}{o1,v1,o2}
	\fmf{photon}{i1,v1}
\end{fmfgraph*}
\end{fmffile}\\
\caption{\label{fig:feynman-diagrams}Feynman diagrams for interactions between weak bosons and fermions in the flavor basis. For reference, $P_L = (1-\gamma_5)/2$, $G_F = \sqrt{2}g^2/8M_W^2$, $\cos\theta_W = M_W/M_Z$ and $I_3^f=\pm1/2$ for the components of a $SU(2)_L$ doublet.}
\end{figure}

In this section we describe the electroweak sector of the standard model Lagrangian, focusing on the Higgs mechanism, the electroweak interactions of fermions, and the flavor structure of the standard model. 
We refer the reader who is less familiar with quantum field theory or who desires a more extensive discussion on the standard model to dedicated books on the subject, such as Refs.~\cite{Zee:2003mt,Quigg:2013ufa}.
The impatient reader who seeks a more applied knowledge may skip this section except for the main results that will be used in the following sections: Eqs.~(\ref{eq:supset-lagrangian}-\ref{eq:currents-2}) describe the weak interactions, see Fig.~\ref{fig:feynman-diagrams} for the Feynman rules; Eq.~(\ref{eq:effective-weak-lagrangian}) is the effective weak Lagrangian after integrating out the weak gauge bosons; and Eq.~\eqref{eq:pmns} parametrizes the mixing in the neutrino sector.

The standard model gauge symmetry is $SU(3)_c\times SU(2)_L\times U(1)_Y$. 
Its particle content is defined as in Table~\ref{tab:sm-fields}.
\begin{table}[t]
\tbl{Standard model fermions and scalar fields. See Eq.~\eqref{eq:higgs-field} and text for the definition of the components of the Higgs field.  \label{tab:sm-fields}}
{\centering
\begin{tabular}{@{}ccccl@{}}
\hline \hline
  Field  &  $SU(3)_c$  &  $SU(2)_L$  &  $U(1)_Y$  &  Description  \\ \hline\hline
  $Q_L=(u_L\,\,d_L)^T$  &  $\boldsymbol{3}$  &  $\boldsymbol{2}$  &  $+1/6$  &  left-handed quarks  \\ \hline
  $u_R$  &  $\boldsymbol{3}$   &  $\boldsymbol{1}$  &  $+2/3$  &  right-handed up-type quarks  \\ \hline
  $d_R$  &  $\boldsymbol{3}$   &  $\boldsymbol{1}$  &  $-1/3$  &   right-handed down-type quarks \\ \hline
  $L=(\nu_L\,\,\ell_L)^T$  &  $\boldsymbol{1}$   &  $\boldsymbol{2}$  &  $-1/2$  &   left-handed leptons \\ \hline
  $\ell_R$  &  $\boldsymbol{1}$   &  $\boldsymbol{1}$  &  $-1$     &   right-handed charged lepton \\ \hline
  $H$  & $\boldsymbol{1}$   &  $\boldsymbol{2}$  &  $+1/2$  &  Higgs scalar  \\ \hline
\end{tabular}}
\end{table}
It is useful to divide the standard model Lagrangian logically into a few parts,
\begin{equation}\label{eq:full-lagrangian}
  \mathcal{L} = \mathcal{L}_{\rm kin}^{\rm gau} + \mathcal{L}_{\rm kin}^{H} + \mathcal{L}_{\rm kin}^{\rm fer}  + \mathcal{L}_{\rm Yuk} - \mathcal{L}_{\rm pot},
\end{equation}
which will be defined later.
We will discuss some of these parts that relate directly to weak interactions and neutrino mixing in more detail. 
We will not discuss quantum chromodynamics, a topic which is well covered elsewhere, e.g. in Refs.~\cite{Ellis:1991qj,Campbell:2017hsr}.

A crucial feature of the standard model is the mechanism of electroweak symmetry breaking. 
This is encoded in the Higgs potential
\begin{equation}\label{eq:higgs-lagrangian}
  \mathcal{L}_{\rm pot} = 
  -M_0^2 H^\dagger H + \frac\lambda2 (H^\dagger H)^2.
\end{equation}
The Higgs field can be written as
\begin{equation}\label{eq:higgs-field}
  H = \frac{1}{\sqrt{2}}\left(
\begin{array}{c} \sqrt{2}G^+_W \\  v+h+iG_Z
\end{array}\right),
\end{equation}
where $G^\pm_W$ and $G_Z$ are the Goldstone modes that will be ``eaten'' by the $W^\pm$ and $Z$ boson giving them masses, as we will see later.
\begin{figure}[t]
\centering
\includegraphics[width=0.45\textwidth]{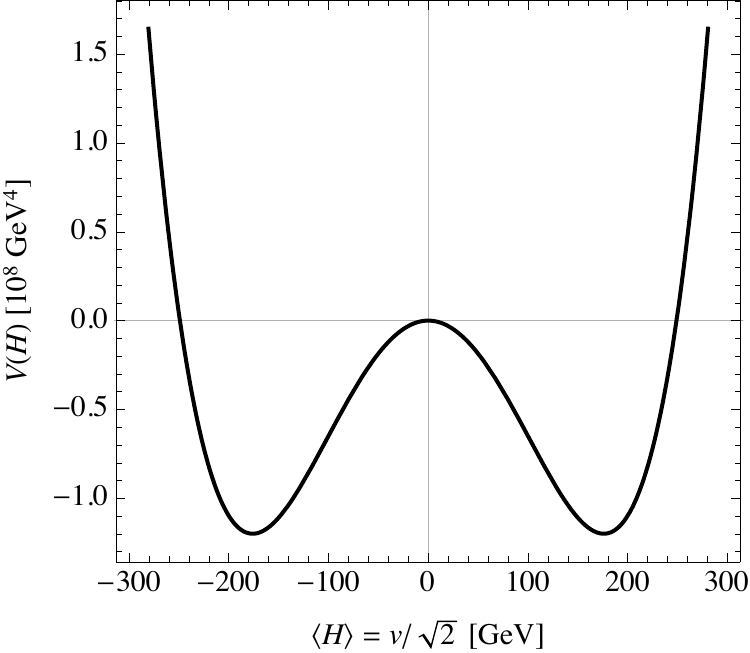}
\caption{\label{fig:higgs-potential}The Higgs potential.}
\end{figure}
For concreteness, we will work on the unitary gauge in which the Goldstone bosons are set to zero, $G_W^\pm,G_Z\to0$.
The vacuum expectation value, or vev for short, $\langle H\rangle=v/\sqrt{2}$ which minimizes the scalar potential shown in Fig.~\ref{fig:higgs-potential}, and the physical Higgs mass are found to be
\begin{equation}
  v^2 = \frac{2M_0^2}{\lambda},\qquad M_h^2 = 2M_0^2.
\end{equation}
Phenomenologically, $M_h\simeq125$~GeV and $v = 1/\sqrt{2}G_F\simeq246$~GeV implies $M_0\simeq88$~GeV and $\lambda\simeq1/4$, approximately.
As the Higgs has a non-zero vev, its kinetic term in Eq.~\eqref{eq:full-lagrangian} gives rise to the masses of the $W^\pm$ and $Z$ gauge bosons.
This can be obtained by expanding the kinetic term of the Higgs, $\mathcal{L}_{\rm kin}^H$. 

Before expanding the kinetic term, it is useful to define our notation for the covariant derivative for a fermions or scalar field $\Psi$ in representations $T^a_c$ and $T^a_L$ of $SU(3)_c$ and $SU(2)_L$, respectively, and with hypercharge $Y$:
\begin{equation}
  D_\mu \Psi = (\partial_\mu + i g_s G_\mu^a T^a_c  + i g W_\mu^a T^a_L + i g' Y B_\mu)\Psi,
\end{equation}
where $G^a_\mu$, $W^a_\mu$ and $B_\mu$ are the gauge fields of the SM gauge group defined above, before spontaneous electroweak symmetry breaking. 
Taking the standard model fermions as an example, the color triplet and weak doublet representations $\{T^a_c\}$ and $\{T^a_L\}$ would be given by the eight Gell-Mann and the three Pauli matrices, respectively.
Note that in our notation, the Gell-mann--Nishijima formula which relates the isospin and hypercharge to the electric charge of particles is  
\begin{equation}
  Q = I_3 + Y,
\end{equation}
where $Q$ is the electric charge and $I_3$ is the isospin (for example $+1/2$ for left-handed up quarks).

The terms that give rise to the gauge boson masses are found in the Higgs kinetic term,
\begin{equation}
  \mathcal{L}_{\rm kin}^H = |D_\mu H|^2 = \left|\left( \partial_\mu  + i \frac{g}{2} W_\mu^a \tau^a + i \frac{g'}{2} B_\mu \right)H \right|^2,
\end{equation}
where $\tau^a$ denotes Pauli matrices.
Expanding this term will yield the physical Higgs boson kinetic term, several 3-particle and 4-particle interactions, and the $W$ and $Z$ masses.
Here, we are interested in the gauge boson mass terms, so we will focus on those only.

We can obtain the mass terms by setting $h\to0$ and noticing that $\partial_\mu v=0$:
\begin{equation}\label{eq:gauge-masses}
  \mathcal{L}_{\rm kin}^H\Big|_{h\to0} = \frac{g^2 v^2}{4} W^+_\mu W^{-\mu} +\frac{1}{2}\left(\frac{g^2+g^{\prime 2}}{4}v^2\right) Z_\mu Z^\mu,
\end{equation}
where 
\begin{align}
  W^\pm_\mu & \equiv (W^1_\mu \mp i W^2_\mu)/\sqrt{2}\\
  Z_\mu & \equiv c_W W^3_\mu  - s_W B_\mu\\
  A_\mu& \equiv s_W W^3_\mu + c_W B_\mu,
\end{align}
and we have defined the weak mixing angle $c_W=\cos\theta_W\equiv g/\sqrt{g^2+g^{\prime 2}}$, and $s_W=\sin\theta_W$.
The weak mixing is one of the electroweak precision parameters and it can be used to probe deviations of the standard model by extracting it at different energies and from different observables like parity violation, gauge boson masses and cross sections.

The expressions for the masses of the gauge bosons and their corresponding measurements performed by high energy collider experiments~\cite{Aaboud:2017svj,D0:2013jba,Aaltonen:2012bp,Abbiendi:2000hu,Abreu:2000mh,Acciarri:2000ai,Barate:1999ce} are \begin{align}
  M_W & = \frac{gv}{2}=80.379\pm{0.012}~{\rm GeV},\\  
  M_Z  & = \frac{gv}{2c_W}=91.1876\pm{0.0021}~{\rm GeV}.
\end{align}
The photon remains massless, as the breaking of electroweak symmetry $SU(2)_L\times U(1)_Y$ leaves an unbroken Abelian gauge group corresponding to electromagnetism, $U(1)_{\rm em}$.

Defining the field strength for the gauge boson $X_\mu^a$ in general as
\begin{equation}
  X^a_{\mu\nu} \equiv \partial_\mu X^a_\nu - \partial_\nu X^a_\nu - g_X f^{abc}V^b_\mu V^c_\nu,
\end{equation}
where $f^{abc}$ is the group's structure constant, e.g. the anti-symmetric Levi-Civita symbol $\varepsilon^{abc}$ for $SU(2)_L$, allows us to write the gauge boson kinetic terms
\begin{equation}
  \mathcal{L}_{\rm kin}^{\rm gau} = -\frac14G_{\mu\nu}^a G^{a\mu\nu} -\frac14W_{\mu\nu}^a W^{a\mu\nu} -\frac14B_{\mu\nu}^a B^{a\mu\nu}.
\end{equation}
We will not go in detail here, but after the gauge boson field redefinitions due to electroweak symmetry breaking, $\mathcal{L}_{\rm kin}^{\rm gau}$ gives rise not only to the kinetic terms themselves but also to triple and quartic gauge interactions among gauge bosons.

The fermionic kinetic term $\mathcal{L}_{\rm kin}^{\rm fer}$ is more important for us, as it encodes the electroweak interactions of fermions.
This term can be written as 
\begin{equation}
  \mathcal{L}_{\rm kin}^{\rm fer} = \sum_\psi^{\rm fermions}\overline\psi i \dslash \psi,
\end{equation}
where $\dslash\equiv \gamma^\mu D_\mu$, with $\gamma^\mu$ being the Dirac matrices, and the sum runs through all fermionic fields, see Table~\ref{tab:sm-fields}.
The terms of interest for us will be the fermion interactions with the $W^\pm$ and $Z$ bosons.

It is useful to write the interaction terms between fermions and weak gauge bosons in the following way
\begin{equation}\label{eq:supset-lagrangian}
  \mathcal{L}_{\rm kin}^{\rm fer} \supset \frac{g}{\sqrt2} \left(J^\mu_W W^+_\mu + J^{\mu\dagger}_W W^-_\mu\right) + \frac{g}{\cos\theta_W}J^\mu_Z Z_\mu 
\end{equation}
where we have defined the charged and neutral weak currents 
\begin{align}\label{eq:currents}
  J_W^\mu\equiv &\sum_{\rm gen.} \bar u \gamma^\mu P_L d + \bar\nu\gamma^\mu P_L \ell,   \\
  J_Z^\mu\equiv &\sum_f \bar f \gamma^\mu \left(I_3^f P_L - \sin^2\theta_W Q_f\right)f, \label{eq:currents-2}
\end{align}
where $P_L\equiv(1-\gamma_5)/2$ is the left-handed projector (similarly, $P_R\equiv(1+\gamma_5)/2$ for the right-handed projector). 
The first the sum runs over all generations, that is, $u$-$d$, $c$-$s$, $t$-$b$, $\nu_e$-$e$, $\nu_\mu$-$\mu$, and $\nu_\tau$-$\tau$.
The second sum runs over all fermions, $I_3^f$ and $Q_f$ are the isospin and electric charge of fermion $f$.
Note that we have implicitly defined the flavor basis in Eq.~\eqref{eq:currents}: the flavor basis is the one in which weak interactions are diagonal, that is, couplings like $usW$, $tdW$, and so on are absent.
Since most neutrino interactions that we will be looking into, as well as neutrino matter effects in oscillation phenomenology, happen at a scale much lower than the electroweak gauge boson masses, it is useful to  integrate out the gauge bosons.
In order to do so, we will set their kinetic term to zero and neglect all terms that involve more than two heavy particle  like triple and quartic gauge couplings and gauge-Higgs interactions.
In this case, equations of motion lead us to~\footnote{The currents with the top quark should also be neglected as its mass is larger than the $W$ mass. To avoid unnecessary clutter, we will leave it implicit.}
\begin{equation}
  \frac{\partial \mathcal{L}}{\partial W_\mu^+} = \frac{g}{\sqrt{2}}J_W^\mu + M_W^2W^{-\mu}=0, \qquad 
  \frac{\partial \mathcal{L}}{\partial Z_\mu} = \frac{g}{\cos\theta_W}J_Z^\mu + M_Z^2Z^{\mu}=0.
\end{equation}
Using this equation to replace the gauge boson fields in Eq.~\eqref{eq:supset-lagrangian} leads to
\begin{equation}\label{eq:effective-weak-lagrangian}
  \mathcal{L}_{\rm weak}^{\rm eff} = -2\sqrt{2}G_F (J^\mu_W J_{W\mu}^\dagger+J^\mu_Z J_{Z\mu}),
\end{equation}
where we have defined the Fermi constant $G_F\equiv \sqrt{2}g^2/8M_W^2\simeq1.166\times10^{-5}$~GeV$^{-2}$. 
As we will see later, this effective weak Lagrangian will be very useful when evaluating four fermion interactions and deriving the neutrino matter potential.

The last piece of the standard model Lagrangian we will discuss is the Yukawa sector,
\begin{equation}
  \mathcal{L}_{\rm Yuk}= -\overline{\boldsymbol{Q}}_L  \boldsymbol{Y}_u \tilde{H} \boldsymbol{u}_R  -\overline{\boldsymbol{Q}}_L  \boldsymbol{Y}_d H \boldsymbol{d}_R  -\overline{\boldsymbol{L}}  \boldsymbol{Y}_\ell H \boldsymbol{\ell}_R + {\rm h.c.},
\end{equation}
where $\tilde H \equiv -i \tau^2 H^*$ and the subscripts ``$L$'' and ``$R$'' denote left- and right-handed fields.
Since we already took the weak interactions to be diagonal in flavor space in Eq.~\eqref{eq:currents}, $\boldsymbol{Y}_u$, $\boldsymbol{Y}_d$ and $\boldsymbol{Y}_e$ should be understood as matrices in flavor space and the bold fermionic fields are vectors in flavor space, e.g., $\boldsymbol{\ell_R} = (e_R,\,\mu_R,\,\tau_R)$. 
The phenomenon of mixing, and thus oscillations, is directly related to this mismatch between weak interactions and mass matrices.
To be more precise, let us write $\mathcal{L}_{\rm Yuk}$ more explicitly as
\begin{equation}
  \mathcal{L}_{\rm Yuk}= -\left(\frac{v+h}{\sqrt{2}}\right)\overline{\boldsymbol{u}}_L  \boldsymbol{Y}_u \boldsymbol{u}_R  - \left(\frac{v+h}{\sqrt{2}}\right)\overline{\boldsymbol{d}}_L  \boldsymbol{Y}_d  \boldsymbol{d}_R  - \left(\frac{v+h}{\sqrt{2}}\right)\overline{\boldsymbol{\ell}}_L \boldsymbol{Y}_\ell \boldsymbol{\ell}_R + {\rm h.c.}
\end{equation}
The mass matrices can be identified as $\boldsymbol{M}_u=\boldsymbol{Y}_u v/\sqrt{2}$, $\boldsymbol{M}_d=\boldsymbol{Y}_d v/\sqrt{2}$ and $\boldsymbol{M}_\ell=\boldsymbol{Y}_\ell v/\sqrt{2}$.
These matrices can be diagonalized by biunitary transformations.
To do this, taking up-type quarks as an example, we can redefine the left- and right-handed fields as $\boldsymbol{u}'_L = V_L^u \boldsymbol{u}_L$, $\boldsymbol{u}'_R = V_R^u \boldsymbol{u}_R$, such that the primed fields define the mass basis as the one in which fermion masses are diagonal, that is,  $\boldsymbol{M}_u^{\text{diag}}=V^{u\dagger}_L \boldsymbol{M}_u V^u_R$. 
A similar procedure can be put forward for down quarks and charged leptons

It follows from this that, since the Higgs vev is the only source of fermion masses,  the Higgs couplings to fermions are also diagonal in the mass basis.
If we turn our attention to the weak charged current in Eq.~\eqref{eq:currents}, we see that weak interactions are not diagonal in the mass basis.
For example, the quark charged current becomes
\begin{equation}
    J_W^\mu\supset \sum_{\rm flavors} \bar{u}'_L \left(V^{u\dagger}_L V^{d}_L\right) \gamma^\mu P_L d'_L,
\end{equation}
where $V^{u\dagger}_L V^{d}_L \equiv V_{\rm CKM}$ is the Cabibbo--Kobayashi--Maskawa mixing matrix~\cite{Cabibbo:1963yz,Kobayashi:1973fv}.
It is curious that the rotations of the right-handed fields have absolutely no physical consequence in the standard model.
A general $3\times 3$ matrix has 18 free parameters. 
Unitarity of $V_{\rm CKM}$, reduces the number of free parameters to 9.
Finally, the six left-handed up- and down-type quark fields can be rephased independently, though a common phase does not affect $V_{\rm CKM}$, and therefore the number of free parameters can be further reduced to 4.
The CKM matrix can be parametrized by three mixing angles and one complex phase, which encodes the violation of charge-parity ($CP$) symmetry.

If we try to follow the same procedure for leptons, we immediately encounter a problem.
Strictly speaking, neutrinos are massless in the standard model, and thus the left-handed neutrino field can always be redefined to absorb any mixing in the leptonic sector.
Nevertheless, the discovery of neutrino oscillations implies that neutrinos  have a nonzero mass~\cite{Kajita:2016cak,McDonald:2016ixn}.
The obvious course of action to implement neutrino masses in the standard model would be to repeat the Higgs mechanism that gives rise to quark masses.
If that would be the case, neutrinos would be Dirac fermions, just like charged leptons and quarks.
The Pontecorvo-Maki-Nakagawa-Sakata (PMNS) matrix $U$~\cite{Pontecorvo:1957qd,Maki:1962mu}, that parametrizes leptonic mixing, would have three mixing angles and one $CP$ violating phase.

Nevertheless, if right-handed neutrinos are included in the Lagrangian, it turns out that these fields are singlets of the standard model, allowing for a bare mass term $M_R\bar\nu_R^c \nu_R$, where the superscript ``$c$'' denotes charge conjugation.
This would give rise to Majorana masses for neutrinos, leading to a completely novel phenomenology with respect to charged fermions.
If the light, active neutrinos are Majorana particles, then $\nu=e^{i\theta}\nu^c$, where $\theta$ is an unphysical phase.
In this scenario, the freedom of rephasing the left-handed neutrino fields would not be there: a rephasing of $\nu_L$ would require a rephasing of $\nu_R$ which in turn would modify the $\nu_R$ bare mass term.
The PMNS matrix would then have three mixing angles and three $CP$ phases. 
The most widely used notation for the PMNS matrix is the following, 
\begin{equation}\label{eq:pmns}
  U = 
\left(
\begin{array}{ccc}
 c_{12}c_{13} & s_{12}c_{13}  & s_{13}e^{-i\delta}  \\
 -s_{12}c_{23}-c_{12}s_{13}s_{23}e^{i\delta}  & c_{12}c_{23}-s_{12}s_{13}s_{23}e^{i\delta}  &  c_{13}s_{23} \\
 s_{12}s_{23}-c_{12}s_{13}c_{23}e^{i\delta}  & -c_{12}s_{23}-s_{12}s_{13}c_{23}e^{i\delta}  &   c_{13}c_{23}
\end{array}
\right)
\cdot
\left(
\begin{array}{ccc}
 e^{i\eta_1}  & 0 & 0  \\
 0  & e^{i\eta_2} & 0  \\
 0  & 0 & 1
\end{array}
\right),
\end{equation}
where $s_{ij}\equiv \sin\theta_{ij}$ and $c_{ij}\equiv\cos\theta_{ij}$, $\delta$ is the so-called Dirac $CP$ phase and $\eta_{1,2}$ are two Majorana phases. 
If neutrinos are Dirac, it suffices to put $\eta_{1,2}=0$ and we recover the most general Dirac mixing matrix.
In fact, there are several other ways of generating neutrino masses besides adding a right-handed neutrino field to the standard model Lagrangian.
The mechanism of neutrino mass is one of the open problems of the standard model, and we will further discuss it by the end of this chapter.

For reference, it is also useful to define some of the Feynman rules for the weak interactions of the standard model.
As a mnemonic, from the Lagrangian to a Feynman rule, one simply adds an $i$ factor to the coefficient in front of the term.
The Feynman rules for $Z$ and $W$ interactions with fermions are given in Fig.~\ref{fig:feynman-diagrams}.

\section{Neutrino interactions}\label{sec:neutrino-interactions}
Neutrinos can be produced by a vast range of processes which populate several orders of magnitude in energy. 
For example, via nuclear fusion reactions the Sun produces electron neutrinos in the 100~keV to 15 MeV energy window, while nuclear reactors produce electron antineutrinos below about 10 MeV via  fission of radioactive isotopes such as $^{235}$U and $^{239}$Pu.
On the other hand, neutrinos produced by proton beams come mostly from the decay of high energy pions and thus are dominantly of muon flavor.
These neutrinos are typically found to have energies from 100~MeV to tens of GeVs or higher.
Atmospheric neutrinos are also come from  the decay of mesons produced in Earth's atmosphere by cosmic rays, and can reach energies of hundreds of TeVs.
Finally, cosmogenic neutrinos have been observed to populate energies up to and above the PeV scale.
In view of this multitude of energies and their corresponding flavor content, it is important to understand the many ways neutrinos can interact with matter and how such interactions can be leveraged to improve our understanding of neutrino properties.
In the following, we will describe the most common and simple ways neutrinos can interact with matter, providing expressions for cross sections wherever appropriate.

\subsection{Neutrino-electron scattering}

\begin{figure}[t]
\centering
 \parbox{1.62in}{	\centering
 	\begin{fmffile}{diagram_nue}
	\begin{fmfgraph*}(100, 65)
		\fmfleft{i1,i2}
		\fmfright{o1,o2}
		\fmflabel{$e^-$}{i1}
		\fmflabel{$\nu_\alpha$}{i2}
		\fmflabel{$e^-$}{o1}
		\fmflabel{$\nu_\alpha$}{o2}
		\fmf{fermion}{i1,v1,o1}
		\fmf{fermion}{i2,v2,o2}
		\fmf{photon,label=$Z$}{v1,v2}
	\end{fmfgraph*}
	\end{fmffile}\vspace{0.5cm}
 \figsubcap{1} \label{fig:nue-diagrams-a}}
 \parbox{1.62in}{	\centering
	\begin{fmffile}{diagram_nue2}
	\begin{fmfgraph*}(100, 65)
		\fmfleft{i1,i2}
		\fmfright{o1,o2}
		\fmflabel{$e^-$}{i1}
		\fmflabel{$\nu_e$}{i2}
		\fmflabel{$\nu_e$}{o1}
		\fmflabel{$e^-$}{o2}
		\fmf{fermion}{i1,v1,o1}
		\fmf{fermion}{i2,v2,o2}
		\fmf{photon,label=$W$}{v1,v2}
	\end{fmfgraph*}
	\end{fmffile}\vspace{0.5cm}
 \figsubcap{2} \label{fig:nue-diagrams-b}}
 \parbox{1.62in}{	\centering
	\begin{fmffile}{diagram_nue3}
	\begin{fmfgraph*}(100, 65)
		\fmfleft{i1,i2}
		\fmfright{o1,o2}
		\fmflabel{$e^-$}{i1}
		\fmflabel{$\bar\nu_e$}{i2}
		\fmflabel{$\bar\nu_e$}{o1}
		\fmflabel{$e^-$}{o2}
		\fmf{fermion}{i1,v1,i2}
		\fmf{fermion}{o1,v2,o2}
		\fmf{photon,label=$W$}{v1,v2}
	\end{fmfgraph*}
	\end{fmffile}\vspace{0.5cm}
 \figsubcap{3} \label{fig:nue-diagrams-c}}
\caption{\label{fig:nue-diagrams}Feynman diagrams for neutrino electron scattering.}
\end{figure}
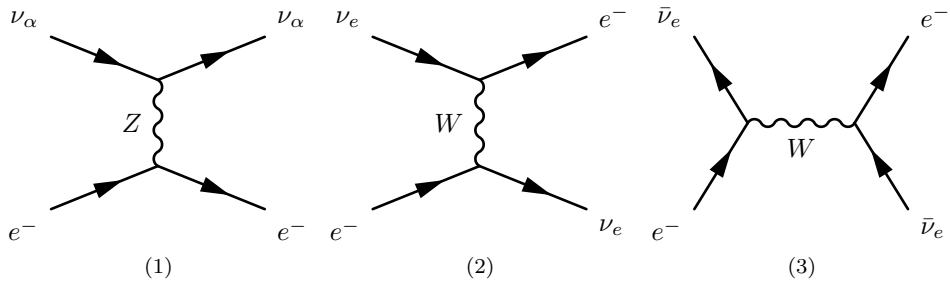

Let us start with a simple, fully electroweak process: neutrino-electron scattering.
Depending on the neutrino flavor this process receive contributions from neutral current and charged current amplitudes, see Fig.~\ref{fig:nue-diagrams}.
For example, in the case of electron neutrinos scattering off electrons, the neutral current amplitude is given by the diagram in Fig.~\ref{fig:nue-diagrams-a} and can be written as
\begin{equation}
  \mathcal{A}_{\rm NC} = i\frac{g^2}{4c_W^2}  \frac{g^{\alpha \beta}-q^\alpha q^\beta/M_Z^2}{M_Z^2-q^2-i\Gamma_ZM_Z}\left[\bar u_\nu(p_3)\gamma_\alpha P_L u_\nu(p_1)\right] \left[\bar u_e(p_4)\gamma_\beta (g_L P_L+g_R P_R) u_e(p_2)\right],\nonumber
\end{equation}
where $g_L=I_3^e-\sin^2\theta_W Q_e = -1/2 + \sin^2\theta_W$ and $g_R=-\sin^2\theta_W Q_e = \sin^2\theta_W$, $\Gamma_Z$ is the $Z$ boson width and $q=p_1-p_2$ is the momentum carried by the virtual $Z$ boson.
The corresponding charged current amplitude in Fig.~\ref{fig:nue-diagrams-b} is 
\begin{align}
  \mathcal{A}_{\rm CC} &= i\frac{g^2}{2} \frac{g^{\alpha \beta}-q^\alpha q^\beta/M_W^2}{M_W^2-q^2-i\Gamma_WM_W} \left[\bar u_e(p_3)\gamma_\alpha P_L u_\nu(p_1)\right] 
  						\left[\bar u_\nu(p_4)\gamma_\beta P_L u_e(p_2)\right]\nonumber\\
			   	   & = i\frac{g^2}{2} \frac{g^{\alpha \beta}-q^\alpha q^\beta/M_W^2}{M_W^2-q^2-i\Gamma_WM_W} \left[\bar u_e(p_3)\gamma_\alpha P_L u_e(p_2)\right] 
  						\left[\bar u_\nu(p_4)\gamma_\beta P_L u_\nu(p_1)\right],\nonumber
\end{align}
where a Fierz identity was used to obtain the second line.
Except for high energy cosmogenic neutrinos, one can usually neglect $q^2$ and the width dependence in the propagator of both amplitudes.
The differential cross section for any neutrino flavor, neglecting these terms, can be written as
\begin{equation}\label{eq:neutrino-electron-scattering}
  \frac{d\sigma_{\nu_\alpha e}}{dE_r} = \frac{2 G_F^2 m_e}{\pi}\left[g_1^2+g_2^2\left(1-\frac{E_r}{E_\nu}\right)^2-g_1g_2\frac{m_e E_r}{E_\nu^2}\right],
\end{equation}
where $E_r$ is the electron recoil energy and $g_{1,2}$ depend on the neutrino flavor, namely,
\begin{align}
  &\nu_e:\qquad ~~\,g_1=1/2+s_W^2\quad {\rm and}\quad g_2 = s_W^2 \\
  &\nu_{\mu,\tau}:\qquad g_1=-1/2+s_W^2\quad {\rm and}\quad g_2 = s_W^2,
\end{align}
and $g_1\leftrightarrow g_2$ to obtain the corresponding antineutrino-electron scattering cross sections.

The total cross section can be obtained by integrating Eq.~\eqref{eq:neutrino-electron-scattering} yielding, to first order in $m_e/E_\nu$,
\begin{equation}\label{eq:neutrino-electron-xsec}
  \sigma(\nu_\alpha e) \simeq \frac{2 G_F^2 m_e E_\nu}{3\pi}(3g_1^2+g_2^2) =  (5.7\times 10^{-44}) (3g_1^2+g_2^2) \left(\frac{E_\nu}{10~{\rm MeV}}\right) ~{\rm cm}^2.
\end{equation}
For example, in the case of $\nu_e$-$e$ scattering we obtain $\sigma(\nu_e e)\simeq 9.5\times 10^{-44}(E_\nu/10~{\rm MeV})~{\rm cm^2}$.
Despite being small, the neutrino-electron scattering cross section is one of the main processes that allows for the study of solar neutrinos in large detectors like Super-Kamiokande.
An important aspect of this process for neutrinos with energies much above the electron mass is the kinematical limit
\begin{equation}
  E_r \theta_e^2 \lesssim 2m_e,
\end{equation}
where $\theta_e$ is the angle between the incoming neutrino and the outgoing electron. 
For high energy recoils, the outgoing electron is very forward.
When studying neutrino-electron scattering, experiments may use this kinematical limit to suppress large backgrounds such as  neutrino-nucleus quasi-elastic interactions.
This indeed was the case for the CHARM-II experiment which measured the vector and axial couplings of electrons to the $Z$ via neutrino-electron scattering~\cite{Vilain:1994qy}.

\begin{figure}[t]
\centering
\includegraphics[width=0.65\linewidth]{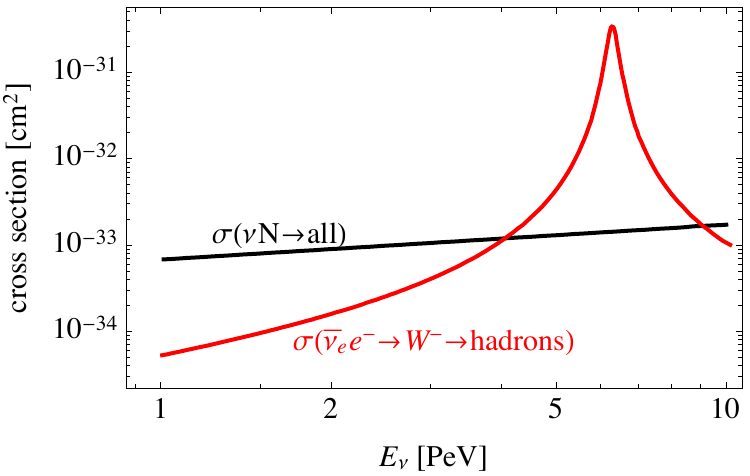}\hspace{1cm}
\caption{Ultrahigh energy neutrino cross sections for the deep inelastic (black) and Glashow resonance (red) processes. Figure adapted from Ref.~\cite{Barger:2014iua}.
\label{fig:glashow}}
\end{figure}

While we have only discussed neutrino-electron scattering at momentum transfers well below the electroweak gauge boson masses, there is one process that takes place at very high energies which deserves attention: the Glashow resonance~\cite{Glashow:1960zz}.
The spectrum of cosmogenic neutrinos extend beyond PeV energies, for which the center of mass energy in neutrino-electron scattering can reach the $W$ boson mass,
\begin{equation}
  \sqrt{s}=\sqrt{2 E_\nu m_e} = \left(\frac{E_\nu}{6.3~{\rm PeV}}\right)^{1/2} M_W.
\end{equation}
The total resonant cross sections for $\bar\nu_e e^-$ scattering can be written as~\cite{Barger:2014iua},
\begin{equation}
  \sigma_{\rm res}^{\bar\nu_e e^-}(s) = 24\pi \Gamma_W^2\frac{s/M_W^2}{(s-M_W^2)^2+M_W^2 \Gamma_W^2}{\rm Br}(W^-\to\bar\nu_e e^-),
\end{equation}
where the last term denotes the branching ratio $W^-\to\bar\nu_e e^-$ found experimentally to be $10.71\pm0.16\%$~\cite{Zyla:2020zbs}.
At $s=M_W^2$ the cross section becomes
\begin{equation}
  \sigma_{\rm res}^{\bar\nu_e e^-}(s=M_W^2) = 4.8\times 10^{-31}~{\rm cm}^2.
\end{equation}
At its peak ($s=M_W^2$), the cross section is much larger than what one would expect from a four fermion interaction, and it is even larger than the neutrino-nucleon deep inelastic scattering cross section at those energies, see Fig.~\ref{fig:glashow}.
Moreover, since it is flavor dependent, it can be used by neutrino telescopes such as IceCube~\cite{Aartsen:2017mau} to provide a statistical handle on the determination of the flavor of cosmogenic neutrinos.

\subsection{Coherent elastic neutrino-nucleus scattering}
Proposed in 1973~\cite{Freedman:1973yd}, the process of coherent elastic neutrino-nucleus scattering, or CEvNS for short, has been recently measured by the COHERENT experiment~\cite{Akimov:2017ade}.
In CEvNS, due to the low momentum transfer ($\lesssim10$~MeV) between the neutrino and the nucleus, the neutrino probes the nucleus as a whole, as opposed to its constituent nucleons.
The scattering takes place via $Z$ boson exchange and its cross section is sensitive to the square of the nucleus (vector) weak charge
\begin{equation}
  Q_V\equiv n_n+(1-4s_W^2)n_p,
\end{equation} 
where $n_{n,p}$ are the number of neutrons and protons in the nucleus $\mathcal{N}$.
This is commonly referred to as the coherent enhancement and can significantly boost the cross section compared to what one would expect if the neutrino was scattering off free nucleons at similar momentum transfers.

Starting from the four fermion Lagrangian in Eq.~\eqref{eq:effective-weak-lagrangian}, we can obtain the differential cross section by coherently summing up the vector coupling of all nucleons or, in other words, treating the nucleus $\mathcal{N}$ as a particle with vector coupling equals $Q_V \mathcal{F}(q^2)$, where  $\mathcal{F}(q^2)$ is a form factor that describes the coherence of the scattering.
The cross section reads
\begin{equation}\label{eq:cevns-xsec}
  \frac{d\sigma_{\rm coh}}{dE_r} \simeq \frac{G_F^2 m_\mathcal{N}}{4\pi}\mathcal{F}^2(q^2)Q_V^2\left( 1- \frac{m_{\mathcal{N}}E_r}{2E_\nu^2} \right),
\end{equation}
where $E_r$ is the recoil energy of the nucleus $\mathcal N$.
The total cross section cannot be calculated analytically due to the form factor, but since $\mathcal{F}(q^2)\le 1$ we can derive a bound on the total cross section
$\sigma_{\rm coh} \le G_F^2 Q_V^2 E_\nu^2/4\pi = 4.2\times 10^{-43}Q_V^2(E_\nu/10~{\rm MeV})^2~{\rm cm}^2$.
Note that the form factor typically plays a significant role in the total cross section for neutrino energies above the 30~MeV or so, depending on the nucleus.

As neutral current processes, to leading order CEvNS is a flavor blind process.
Although it is a relatively large cross section, using CEvNS to study terrestrial or astrophysical neutrino sources does have its challenges.
Due to the nature of this scattering most of the incoming neutrino energy is carried away by the outgoing neutrino and thus is undetectable.
The recoiled nucleus carries a small fraction of the incoming neutrino energy.
Kinematics restricts $E_r\lesssim 2E_\nu^2/m_\mathcal{N}$ which is usually below a few keV for most nuclei used in detecting CEvNS events.
The lack of a smoking gun experimental signature and the low deposition of energy makes it very difficult to suppress backgrounds.
The COHERENT experiment has measured CEvNS by deploying CsI~\cite{Akimov:2017ade} and argon~\cite{Akimov:2020pdx} detectors near a stopped pion source at the Spallation Neutron Source (SNS) at Oak Ridge National Laboratory.
Shielding and beam time structure enable COHERENT to reduce backgrounds and observe evidence of CEvNS consistent with the expectations in the standard model ($134\pm22$ CEvNS events observed, compared to $173\pm48$ expected).

\subsection{Quasi-elastic neutrino-nucleon scattering}
Another process which is very important for neutrino measurements is the quasi-elastic scattering of neutrinos off nucleons, namely, $\nu n \to \ell^- p^+$ and $\bar\nu p^+ \to \ell^+ n$ via charge current interactions.
In the regime where the momentum transfer is much smaller than the nucleon mass, $-q^2\ll m_N^2$, neutrinos will not probe the internal structure of the nucleon.
If the neutron and proton were to be elementary particles, the calculation would be very similar to what we have done for neutrino-electron scattering.
Nevertheless, because nucleons are not elementary, we need to parametrize their couplings to the weak gauge bosons by form factors.

To be more concrete, let us take $\bar\nu p^+ \to \ell^+ n$ as an example.
The amplitude for this process, since $-q^2\ll m_N^2\ll M_W^2$, can be written as 
\begin{equation}
  \mathcal{A} = \bra{\ell^+ n} \mathcal{O} \ket{\bar\nu p^+},
\end{equation}
where $\mathcal{O}$ is the four-fermion operator that describes the transition which we assume to be of the form 
\begin{equation}
  \mathcal{O}=2\sqrt{2}G_F \cos\theta_{\rm C} (\bar\nu\gamma_\mu P_L \ell)\left\{\bar n \gamma^\mu \left[\frac12F_V(q^2) + \frac12F_A(q^2)\gamma^5\right] p\right\}.
\end{equation}
$F_V$ and $F_A$ are the vector and axial form factors that depend on the momentum transfer $q^2$ and $\cos\theta_{\rm C}\simeq0.974$ is the cosine of the Cabibbo angle~\cite{Zyla:2020zbs}.
In principle, other form factors (e.g. tensor) could also be present, but we neglect those since their effect on the cross section is subleading.
The amplitude can be cast in the form
\begin{equation}
  \mathcal{A} = 2\sqrt{2}G_F \cos\theta_{\rm C} \left[\bar v_\nu(p_1)\gamma_\mu P_L v_\ell(p_3)\right]
  									\left\{\bar u_n(p_4)\gamma^\mu \frac12 \left[F_V(q^2)+F_A(q^2)\gamma^5\right]u_p(p_2)\right\}.
\end{equation}
For low momentum transfer, $q^2\to0$, the form factors become the vector and axial coupling constants $g_V$ and $g_A$. 
Calculating the total cross section in the lab frame is straightforward and leads to
\begin{equation}\label{eq:neutrino-QE-xsec}
  \sigma_{\rm QE}^{\bar\nu p}=\frac{G_F^2 E_\nu^2}{\pi}\cos^2\theta_{\rm C}\left(g_V^2+3g_A^2\right)\simeq 9.4\times 10^{-42}\left(\frac{E_\nu}{10~{\rm MeV}}\right)^2~{\rm cm}^2,
\end{equation}
where we have assumed $g_V=1$ (conservation of vector current) and the determination of the axial coupling from $\beta$ decay measurements~\cite{Zyla:2020zbs} yields $g_A=-1.2756$. 

By comparing Eqs.~\eqref{eq:neutrino-electron-xsec} and \eqref{eq:neutrino-QE-xsec} we observe that this cross section is substantially larger than neutrino-electron scattering and may yield much higher statistics in neutrino experiments.
While the outgoing charged lepton in quasi-elastic scattering allows for flavor identification, the non-negligible mass of muons and taus, compared to nucleons, imposes a minimum energy threshold for the reaction to occur, 
\begin{equation}\label{eq:threshold}
  E_\nu^{\rm min}\simeq m_\ell + \Delta +\frac{m_\ell^2}{2m_N},
\end{equation}
where $\Delta\equiv m_N^{\prime}-m_N$ and $N,N'$ denote the struck and outgoing nucleons, respectively. 
For reference, $m_n-m_p\simeq 1.3$~MeV. 
For the inverse beta decay process, $\bar\nu_e p^+\to e^+ n$, the threshold of 1.8~MeV plays an important role in the detection of reactor and geoneutrinos. 
For quasi-elastic scattering of muon neutrinos, the threshold is approximately the muon mass.
Tau neutrino production requires a much larger incoming neutrino energy of about $3.5$~GeV, as well as larger momentum transfers.

Large detectors with considerable amounts of free protons, such as water Cherenkov detectors like Super-Kamiokande~\cite{Ikeda:2007sa} and scintillator detectors such as Daya Bay~\cite{Guo:2007ug}, RENO~\cite{Ahn:2010vy} and NOvA~\cite{NOvA:2020dll}, may take advantage of the $\bar\nu_e p^+\to e^+ n$ process when considering astrophysical measurements such as supernova explosions. 
In some detectors, the outgoing neutron can be captured by nuclei (e.g. gadolinium) inducing a photon emission, leading to a richer experimental signature that allows for background reduction.
Finally, this process may also occur with bound nucleons. 
In this case, the calculation of the cross section needs to take into account nuclear effects and the energy thresholds in general differ significantly from the free proton one.
Of particular interest are the quasi-elastic transitions of $\nu_e$-$^2{\rm H}$, $\nu_e$-${\rm Ga}$ and $\nu_e$-${\rm Cl}$, which were used to study solar neutrinos~\cite{Davis:1964hf,Hirata:1989zj,Hirata:1991ub,Abazov:1991rx,Anselmann:1992um,Anselmann:1993mh,Anselmann:1994cf,Abdurashitov:1994bc,Aharmim:2011vm}, as well as $\nu_e$-${\rm C}$, $\nu_e$-${\rm Ar}$, $\nu_e$-${\rm Pb}$, and $\nu_e$-${\rm Fe}$ which can be used to detect supernova neutrinos~\cite{NOvA:2020dll,Athar:2006yb,Zuber:2015ita,Abi:2020lpk}.
The quasi-elastic cross section is also important for beam neutrino experiments.

\subsection{Deep Inelastic Scattering}
At very high energies, neutrinos can probe the components of nucleons.
In contrast to the quasi-elastic scattering, where neutrinos induce a transition between nucleons, in the deep inelastic scattering (DIS) regime, $-q^2 \gtrsim m_p^2$, the neutrino scatters off partons inside the proton, which then may shower to further quarks and gluons, and finally hadronize. 
Although a detailed treatment of the high energy neutrino-hadron interactions is beyond the purpose of this chapter, it is useful to show how a calculation would be performed in the DIS regime using the parton model.
For this end, we will calculate $\nu p^+ \to \ell^- X$, where $X$ represents anything else coming out of the interaction.

The central idea of the parton model is that the the scattering on nucleons is given by the incoherent sum of the scattering on its partons.
We start by denoting the probability distribution function of a parton $i$ (e.g., an up quark or a gluon) carrying a fraction $x$ of the proton momentum by $f_i(x)$.
Since $f_i(x)$ is a probability distribution, it follows certain sum rules, namely
\begin{align}
  &\sum_i^{\rm partons}\int_0^1 dx\, x f_i(x) = 1 &\text{(total momentum),} \\
  &\int_0^1 dx \left(\frac23[f_u(x)-f_{\bar u}(x)] - \frac13[f_d(x)-f_{\bar d}(x)] \right)= 1 & \text{(proton  charge),} \\
  &\int_0^1 dx \left(\frac13[f_u(x)-f_{\bar u}(x)] - \frac23[f_d(x)-f_{\bar d}(x)]\right) = 0 & \text{(neutron  charge).} 
\end{align}

For the total cross section,  we can write
\begin{equation}
  \sigma(\nu p^+ \to \ell^- X)=\sum_i^{\rm partons}\int dx \frac{d\sigma_i}{dx}=\sum_i^{\rm partons}\int dx\, \sigma_i(x) f_i(x),
\end{equation}
where the index $\sigma_i(x)$ is the cross section for a neutrino scattering off parton $i$ carrying a momentum fraction $x$ of the proton.
For simplicity we will only consider first family partons: $u$, $d$, $\bar u$ and $\bar d$.
Typically, the following variables are defined
\begin{align}
  Q^2 &\equiv -q^2 = (p_1-p_3)^2,\\
  x &\equiv \frac{Q^2}{2M\nu},\\
  \nu &\equiv E-E' \equiv yE,
\end{align}
where $p_1$ and $p_3$ and the momenta of the incoming neutrino and outgoing charged letpon, and $x$ is the Bjorken scaling variable which describes the momentum fraction carried by the parton, $\nu$ is the energy transfer given by the difference between the incoming neutrino energy $E$ and outgoing lepton energy $E'$, and $y$ is called the inelasticity parameter.
The nucleon mass is denoted by $M$. 

To calculate the DIS cross section in the lab frame (that is, neutrinos scattering off nucleons at rest) it suffices to do
\begin{equation}
  \frac{d^2\sigma}{dxdy} = 2MEx \sum_i^{\rm partons}\frac{1}{64\pi}\frac{1}{M^2 E^2  x^2}\overline{|\mathcal{A}_i(x,y)|^2}f_i(x),
\end{equation}
where the amplitude $\mathcal{A}_i(x,y)$ is calculated assuming that the parton $i$ carries a fraction $x$ of the momentum of the proton and the bar denotes the spin average of incoming particles. 
If we neglect the masses of all fermions and integrate out the $W$, see Lagrangian in Eq.~\eqref{eq:effective-weak-lagrangian}, the $\nu d\to \ell^- u$ squared amplitude is given by
\begin{align}
  \mathcal{A}_{\nu d} &= 2\sqrt{2}G_F\left(\bar u_\ell \gamma_\alpha P_L u_\nu\right)\left(\bar u_{u}\gamma^\alpha P_L u_{d}\right)\\
  \Rightarrow |\mathcal{A}_{\nu d}|^2 & \simeq 8 G_F^2 {\rm Tr}(\pslash_3 \gamma_\alpha P_L \pslash_1\gamma_\beta P_L)
  									{\rm Tr}(\pslash_4 \gamma^\alpha P_L \pslash_2\gamma^\beta P_L) = 32G_F^2\hat s^2,
\end{align}
where we have defined $\hat s = x s \simeq 2MEx$, the  Mandelstam  $s$ variable of the neutrino-parton system.
A similar calculation can be done for neutrinos interacting with $\bar u$ quarks yielding, $|\mathcal{A}_{\nu \bar u}|^2=32G_F^2\hat u^2 = 32G_F^2 \hat s^2 (1-y)^2$.
For antineutrinos, one would obtain $|\mathcal{A}_{\bar\nu u}|^2=32G_F^2 \hat s^2 (1-y)^2$ and $|\mathcal{A}_{\bar \nu \bar d}|^2 = 32G_F^2 \hat s^2$.

One important aspect of these squared amplitudes is their different $y$ dependence.
This is related to the helicity of the incoming neutrino and the chirality of weak interactions,  which is encoded in the left-handed projector $P_L$ in the $W$ vertex.
Considering only valence quarks, we can see that the antineutrino squared amplitude is suppressed for large inelasticity, since $|\mathcal{A}_{\bar \nu u}|^2\to0$ as $y\to1$. 
Therefore in charged current antineutrino deep inelastic scattering, a larger fraction of the incoming neutrino energy is carried by the outgoing lepton in comparison to the outgoing hadronic system.
This can have important experimental consequences, as in neutrino experiments it is usually much easier to reconstruct the leptonic final state compared to the hadronic one.

Combining all these results and noting that the incoming neutrino is fully polarized and so the average on spins is a 1/2 factor, we obtain the following cross sections
\begin{align}
  \frac{d^2\sigma}{dxdy}(\nu p^+\to\ell^- X) = \frac{G_F^2ME}{\pi} 2x \left[f_d(x) +(1-y)^2f_{\bar u}(x)\right],\\
  \frac{d^2\sigma}{dxdy}(\bar\nu p^+\to\ell^+ X) = \frac{G_F^2ME}{\pi} 2x \left[f_{\bar d}(x) +(1-y)^2f_{u}(x)\right].
\end{align}
Now, consider that neutrinos scatter on an isoscalar target $N$, i.e. nuclei that contain an equal amount of protons and neutrons.
Under isospin symmetry, the $u$ quark pdf in the neutron is equal to the $d$ quark pdf in the proton, and so on. 
Then, the DIS cross section of an isoscalar target can be written as
\begin{align}
  \frac{d^2\sigma}{dxdy}(\nu \mathcal N\to\ell^- X) = \frac{G_F^2ME}{\pi} 2x \left\{f_u(x) + f_d(x) +(1-y)^2[f_{\bar u}(x)+f_{\bar d}(x)]\right\},\\
  \frac{d^2\sigma}{dxdy}(\bar\nu \mathcal N\to\ell^+ X) = \frac{G_F^2ME}{\pi} 2x \left\{(1-y)^2[f_{u}(x)+f_d(x)] + f_{\bar u}(x)+f_{\bar d}(x)\right\}.
\end{align}

Neutrino nucleus deep inelastic scattering is an important process for neutrinos above several GeV.
In particular, atmospheric neutrinos above around 5~GeV will interact mainly through DIS.
For astrophysical neutrinos with energies above 100~TeV, such as those detected at the IceCube Neutrino Observatory, DIS is by far the dominant interaction mode, except at the Glashow resonance energy.

\subsection{Neutrino-nucleus interactions at the GeV scale}

The attentive reader may have noticed that we went from quasi-elastic neutrino-nucleon interactions to deep inelastic scattering, skipping the 100~MeV--few GeV regime.
The reason is not the lack of importance.
In fact, this energy range is one of the most relevant for neutrino oscillation measurements due to the values of the mass splittings and the muon mass, and here is why.
The oscillation phase is given by $1.27\Delta m^2_{ij} [{\rm eV}^2]L[{\rm km}]/E[{\rm GeV}]$.
For an experiment to measure muon neutrinos via charged current interactions, the minimum neutrino energy to overcome the muon mass threshold is about 100~MeV, see Eq.~\eqref{eq:threshold}.
Nevertheless, near the threshold the cross section is suppressed by phase space, so let us consider a neutrino energy of at least a few hundred MeV for the sake of this argument.
The largest mass splitting is about $2.5\times 10^{-3}$~eV$^2$.
If one considers a baseline of at most 2000~km, comparable to the radius of the Earth of 6371~km, the maximum neutrino energy that would lead to an oscillation phase of $\pi/2$ would be about 4~GeV, while shorter baselines would require lower energies.
Matter effects will affect the oscillation frequency, but even then this general argument holds.
Therefore, to be able to observe muon neutrinos undergoing significant oscillations, the energy spectrum of these neutrinos should be within few hundred MeV to several GeV.

\begin{figure}[t]
\centering
\includegraphics[width=0.49\linewidth]{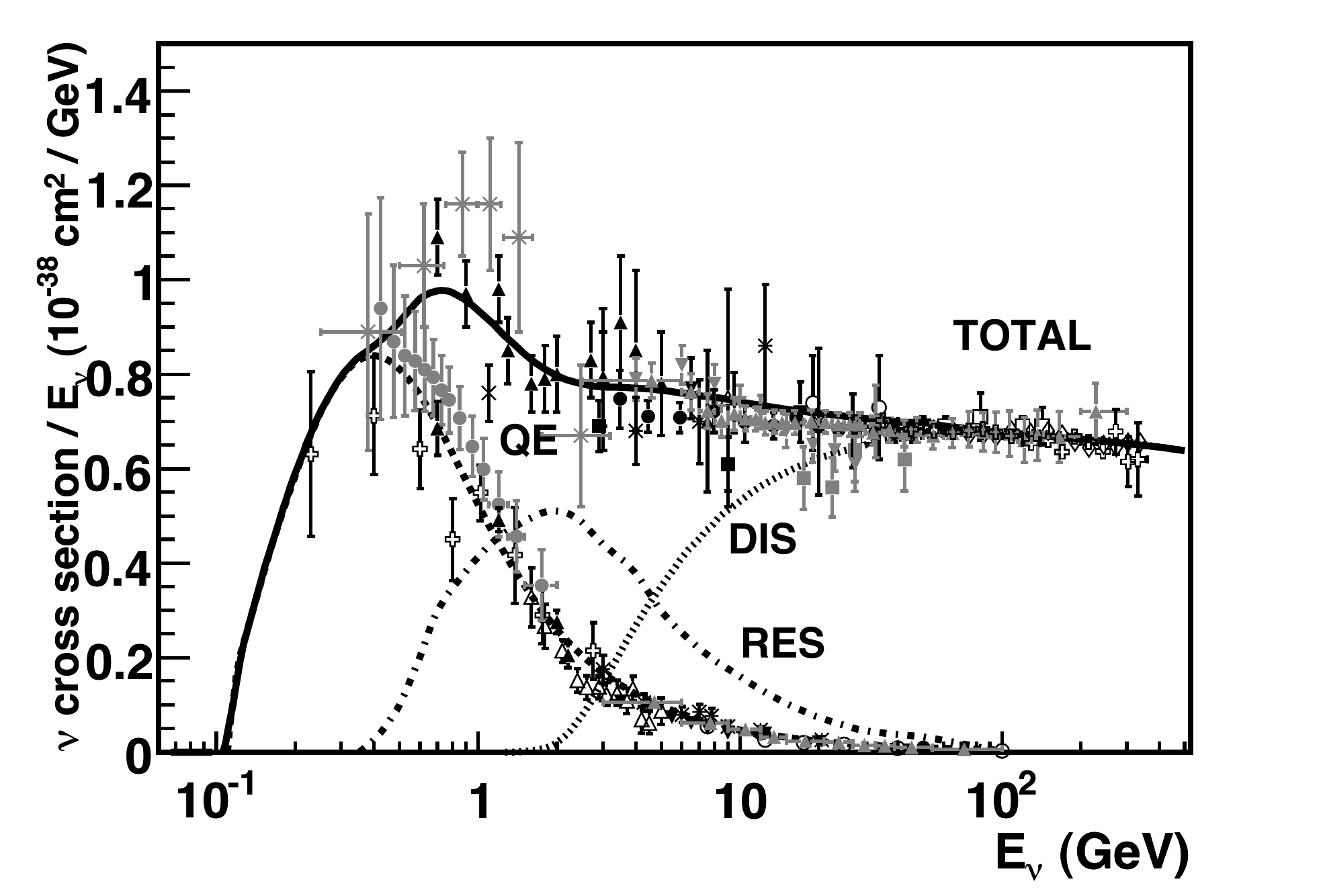}
\includegraphics[width=0.49\linewidth]{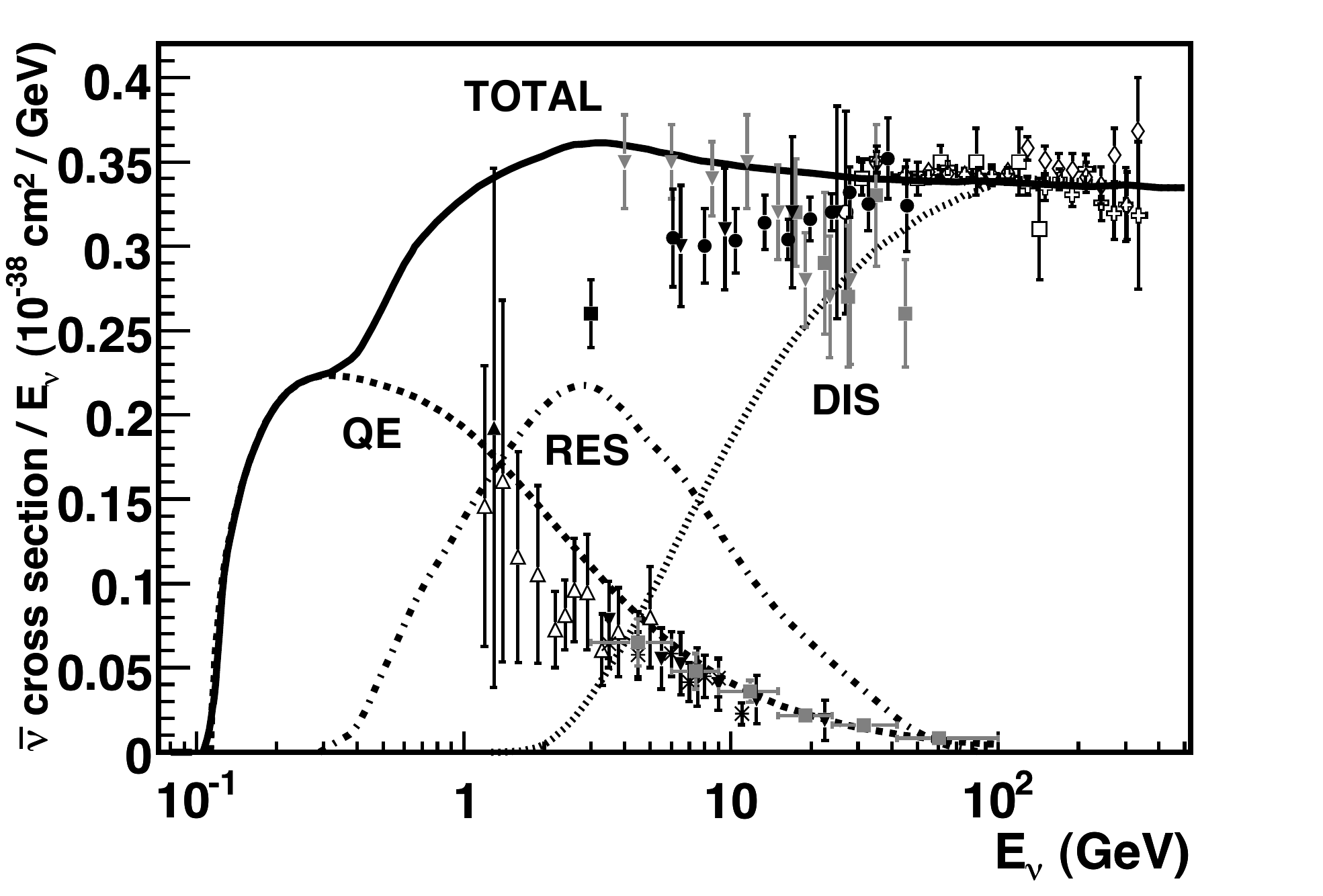}
\caption{Muon neutrino (left) and muon antineutrino (right) total cross section per nucleon. Plot taken from Ref.~\cite{Formaggio:2013kya}.
\label{fig:neutrino-xsec}}
\end{figure}

The difficulty in describing neutrino-nucleus interactions at this energy scale, comes from a complex interplay between nuclear physics, for example the propagation of nucleons or pions throughout the nucleus, and non-perturbative QCD. 
Although for high energy neutrinos, much above 10~GeV or so, one can properly describe neutrino-nucleus interactions by the deep inelastic scattering formalism (with some modifications due to the nuclear medium), as the neutrino energy is lowered the parton model breaks down.
Neutrino interactions may excite resonances such as $\Delta$'s or even interact with two correlated nucleons at the same time.
A  description of the current knowledge of neutrino-nucleus interactions at the GeV scale is certainly beyond the scope of this book, but we can nevertheless look at the experimental data on neutrino-nucleon interaction cross section in Fig.~\ref{fig:neutrino-xsec} for muon neutrinos (left) and muon antineutrinos (right), taken from Ref.~\cite{Formaggio:2013kya} and references therein. 
The lines are theoretical predictions for quasi-elastic (QE), resonant cross section (RES) and deep inelastic scattering (DIS).

%
%


\section{Neutrino oscillations}\label{sec:oscillations}

One may argue that the oscillation of neutrinos is the most unique property of these particles. 
It is a quantum mechanical effect that takes place over macroscopic distances, and it is the only evidence so far that neutrinos have nonzero masses.
In what follows, we will describe the formalism of neutrino oscillations in vacuum and in matter and highlight some special properties of oscillation physics.

\subsection{Neutrino oscillations in vacuum}
It is fairly straightforward to derive the equation that describes the evolution of neutrinos.
For a neutrino field $\psi(x)$, the free particle Dirac equation 
\begin{equation}
  (i\psl -m)\psi(x)=0
\end{equation}
has a plane wave solution 
\begin{equation}
  \psi(x) = u(p)e^{-ip\cdot x},
\end{equation}
where $p_\mu$ and $x_\mu$ are the energy-momentum and space-time four-vectors.
The neutrino dispersion relation $E^2=p^2+m^2$ follows from the Dirac equation and it is trivial that
\begin{equation}\label{eq:1-neutrino-evolution}
  i\frac{\partial}{\partial t}\psi(x) = E\psi(x) \simeq \left(p + \frac{m^2}{2p}\right)\psi(x),
\end{equation}
where a first order expansion on $m/p$ was performed in the last term which is excellent approximation for all practical purposes due the the smallness of neutrino masses (as we will see in Sec.~\ref{sec:questions}, neutrino masses are experimentally constrained to be below a couple electronVolts).
To understand neutrino oscillations, we start by emphasizing that the neutrino flavor eigenstate produced via weak interactions $\ket{\nu_\alpha}$, where $\alpha=e,\mu,\tau$, is a linear superposition of mass eigenstates 
\begin{equation}\label{eq:flavor-mass}
  \ket{\nu_\alpha}=\sum_{i=1}^3 U^*_{\alpha i}\ket{\nu_i},
\end{equation} 
where $\ket{\nu_i}$ have well-defined masses $m_i$ and their evolution follows Eq.~\eqref{eq:1-neutrino-evolution}, plugging in the necessary generation indices. 
Note that the change from mass to flavor bases is consistent with the relation $\psi_\alpha=\sum_{i=1}^3 U_{\alpha i}\psi_i$ for the neutrino fields in the flavor and mass basis, since $\ket{\nu}=\psi^\dagger\ket{0}$. 
$U$ is the PMNS matrix defined in Eq.~\eqref{eq:pmns}.
The Majorana phases do not play any role in oscillation phenomenology and  can be redefined away if neutrinos are Dirac particles.
We will discuss the measurements of oscillation parameters in Sec.~\ref{sec:measurements}.

For clarity, we will use latin indices ``$i,j,\dots$'' when working in the mass basis while greek indices ``$\alpha,\beta,\dots$'' will be reserved for the flavor basis.
We can now identify the neutrino evolution operator in vacuum, i.e. the free Hamiltonian, as $H^0_{ij} = (p + m_i^2/2p)\delta_{ij}$ written in the mass basis (denoted by a superscript ``0'').
Note that the neutrino energy and momentum are almost identical and thus one can interchange $p\leftrightarrow E$ in the definition of the Hamiltonian and in Eq.~\eqref{eq:1-neutrino-evolution}.
Although the evolution of neutrino mass eigenstates is fairly trivial, neutrinos produced and detected via  charged current weak interactions have well defined flavors.
The Hamiltonian in the flavor basis can be obtained as $H=U^\dagger H^0U$.
We can therefore write the probability for a neutrino of a flavor $\alpha$ to oscillate into a neutrino of flavor $\beta$ after it propagates a distance $L=c\,t$ in vacuum as
\begin{align}\label{eq:transition-amplitude}
  P(\nu_\alpha\to\nu_\beta;L)&=|\mathcal{A}_{\alpha\beta}(L)|^2=|\bra{\nu_\beta}e^{-iHL}\ket{\nu_\alpha}|^2\\
  	& = \sum_{i,j,k,l}^{1..n}U_{\alpha i}^*U_{\beta j}U_{\alpha k}U_{\beta l}^* \braket{\nu_j}{\nu_i}\braket{\nu_k}{\nu_l} \exp \left[-i \frac{(m^2_i-m^2_k)L}{2E}\right]\\\label{eq:transition-amplitude-final}
  	& = \sum_{i,j}^{1..n}U_{\alpha i}^*U_{\beta i}U_{\alpha j}U_{\beta j}^* \exp \left(-i \frac{\Delta m^2_{ij}L}{2E}\right),
\end{align}
where the sum runs over all neutrino mass eigenstates (which in the standard three neutrino framework is $n=3$), and $\Delta m^2_{ij}\equiv m^2_i-m^2_j$.
By working out the trigonometric functions and using the unitarity of the PMNS matrix, the oscillation probability can be written in a way that explicitly shows it is a real quantity~\cite{Zyla:2020zbs},
\begin{align}
  P(\nu_\alpha\to\nu_\beta;L)=\delta_{\alpha\beta} &-  4\sum_{i<j}^{1..n}{\rm Re}\left(U_{\alpha i}^*U_{\beta i}U_{\alpha j}U_{\beta j}^*\right) \sin^2\left(\frac{\Delta m^2_{ij}L}{4E}\right) \nonumber\\ \label{eq:probability}
  & + 2\sum_{i<j}^{1..n}{\rm Im}\left(U_{\alpha i}^*U_{\beta i}U_{\alpha j}U_{\beta j}^*\right) \sin^2\left(\frac{\Delta m^2_{ij}L}{2E}\right).
\end{align}

The overall phase, the leading term in Eq.~\eqref{eq:1-neutrino-evolution}, drops out in the oscillation probability, while the dependence on the Majorana phases completely cancels out.
It is important to note that antineutrinos are created by $\psi$, as opposed to $\psi^\dagger$, and thus the equivalent of Eq.~\eqref{eq:flavor-mass} for antineutrinos is $\ket{\bar\nu_\alpha}=\sum_i U_{\alpha i}\ket{\bar\nu_i}$. 
Therefore, $P(\bar\nu_\alpha\to\bar\nu_\beta;L)$ can be obtained by replacing $U\leftrightarrow U^*$ everywhere in Eq.~\eqref{eq:probability}.
Under $CP$ conjugation $\nu_\alpha$ goes to $\bar\nu_\alpha$.
Thus, since conjugating $U$ will flip the sign of all phases, we can understand why they are related to $CP$ violation. 
In the literature,  $\delta$ is typically referred to as the Dirac  or the $CP$ violation phase, even if other phases that could be present in the PMNS matrix also encode $CP$ violation (e.g. the Majorana phases or extra Dirac phases in scenarios with more than three neutrinos).

When comparing the $P(\nu_\alpha\to\nu_\beta;L)$ to $P(\bar\nu_\alpha\to\bar\nu_\beta;L)$ we see that only the last term in Eq.~\eqref{eq:probability} changes, and therefore the first and second terms are $CP$-conserving while the last is $CP$-violating.
Furthermore, we can see that $CP$ violation cannot be directly observed in disappearance oscillation channels, that is, $\alpha=\beta$, as in that case the last term in Eq.~\eqref{eq:probability} vanishes~\footnote{In principle, measuring independently, via the disappearance channels, the absolute value of four entries of the PMNS matrix, or a combination of those, which are not related by unitarity (e.g., $U_{e2}$, $U_{e3}$, $U_{\mu2}$ and $U_{\mu3}$) would be sufficient to provide a determination of $\delta$. It turns out that this is impractical for a number of reasons. 
}.
As a final comment, although we have used $\delta$ to discuss the violation of $CP$ symmetry, how we encode $CP$ violation in phases is parametrization dependent.
A basis independent way of quantifying the amount of $CP$ violation is via the use of the Jarlskog invariant~\cite{Jarlskog:1985ht,Jarlskog:1985cw}, $J\equiv\text{Re}(U_{\mu 3}U_{e 3}^*U_{\mu 2}^*U_{e 2})$, which in our notation is given by $J = s_{23}c_{23}s_{13}c_{13}^2s_{12}c_{12}\sin\delta$.

It turns out that it is sometimes useful to work in a simplified 2-neutrino framework.
In this case, the PMNS matrix is just a simple rotation in two dimensions, without any $CP$ phase, and Eq.~\eqref{eq:probability} greatly simplifies.
For example, the survival or disappearance probability is 
\begin{equation}\label{eq:two-flavor-oscillation}
  P(\nu_\alpha\to\nu_\alpha) = 1 - \sin^2(2\theta)\sin^2\left(\frac{\Delta m^2 L}{4E}\right)\simeq1 - \sin^2(2\theta)\sin^2\left(1.27\frac{\Delta m^2 [{\rm eV}^2] L[{\rm km}]}{E[{\rm GeV}]}\right).
\end{equation}

Although the measurements of  oscillation parameters will be discussed in Sec.~\ref{sec:measurements}, we can anticipate some results so that reader can get an idea of the phenomenology of neutrino oscillations. 
We have measured so far two mass splittings, the solar one as $\Delta m^2_{21}\simeq 7.5\times10^{-5}~{\rm eV}^2$, and the atmospheric splitting $|\Delta m^2_{31}|\simeq 2.5\times10^{-3}~{\rm eV}^2$, as well as the three mixing angles $\theta_{12}\simeq33^\circ$, $\theta_{13}\simeq8^\circ$ and $\theta_{23}\simeq45^\circ$.
Given the largest mass splitting, we can already appreciate why neutrino experiments need such long distances to observe oscillations: a GeV neutrino would need to propagate hundreds of kilometers to develop sizable oscillations!

\begin{figure}[t]
\centering
\includegraphics[width=0.98\linewidth]{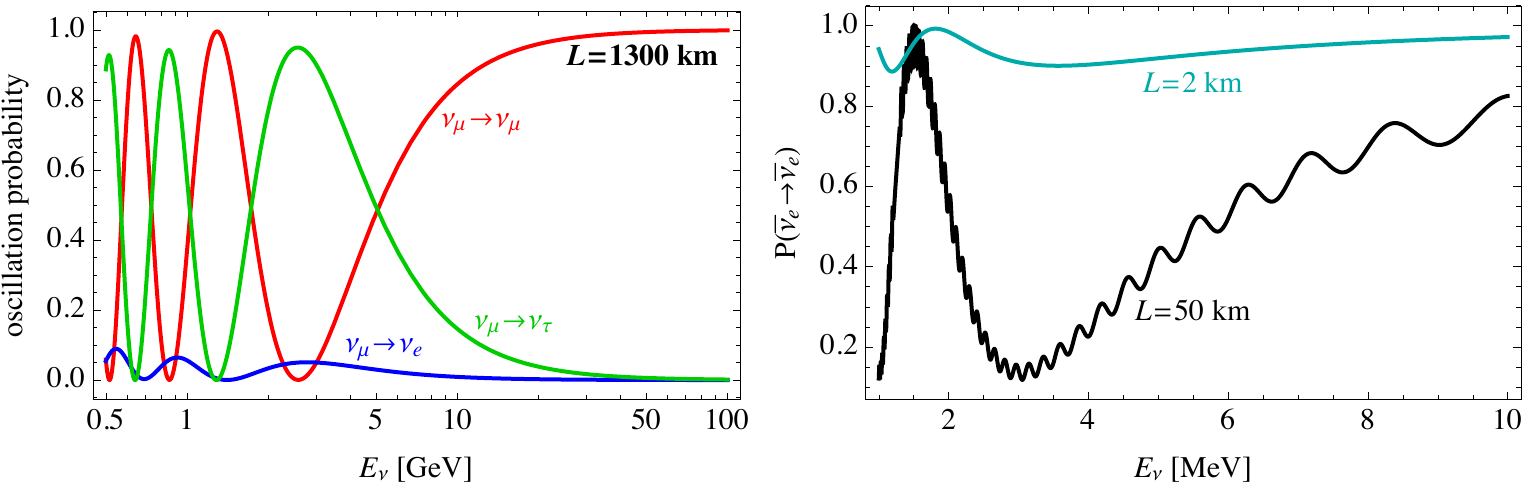}
\caption{Left: Oscillation probabilities at a distance of 1300~km for a beam of muon neutrinos with energies at the GeV scale: $\nu_\mu\to\nu_\mu$ disappearance channel (red) and the $\nu_\mu\to\nu_\tau$ (red) and $\nu_\mu\to\nu_e$ (blue) appearance channels. Right: $\bar\nu_e$ disappearance channel for energies relevant to reactor antineutrinos for two different baselines, 2~km (cyan) and 50~km (black). For simplicity, probabilities were calculated in vacuum.
\label{fig:oscillations}}
\end{figure}

To give a more precise picture, we show in Fig.~\ref{fig:oscillations} three important oscillation channels for the future DUNE experiment (left panel), which would have a baseline of 1300~km and a beam with mostly muon neutrinos, as well as the disappearance oscillation probability for reactor neutrinos (right panel) at two different baselines, 2~km and 50~km.
For simplicity, oscillations were calculated in vacuum.
Focusing on the left panel, muon neutrino disappearance at ``atmospheric baselines'' is driven by the largest mixing angle $\theta_{23}$ and can be quite large (red).
Most of these muon neutrinos oscillate into tau neutrinos (green), while a small fraction goes to electron neutrinos (blue).
The latter effect is driven by $\theta_{13}$ which is relatively small.
In the right panel, we see that $\bar\nu_e\to\bar\nu_e$ disappearance channel for MeV neutrinos  at 2~km is relatively small (cyan), as again it is driven by $\theta_{13}$. 
Nevertheless, as one goes even further away to 50~km (black), the solar oscillation frequency kicks in and the $\bar\nu_e$ disappearance becomes significant, as it is driven by $\theta_{12}$.
The small wiggles in the black line are due to the atmospheric splitting.

This simple example of oscillation phenomenology serves to show how rich neutrino phenomenology can be and how challenging it may be to experimentally isolate and understand all aspects of oscillations. 
We will discuss oscillation phenomenology and corresponding measurements in Sec.~\ref{sec:measurements}, but before that let us consider how the presence of matter may  affect neutrino propagation and oscillation physics.

\subsection{Neutrino oscillations in matter}\label{sec:oscillations-in-matter}
The presence of matter sources an effective potential that changes the neutrino dispersion relation in dense media~\cite{Wolfenstein:1977ue,Mikheev:1986gs}. 
This is similar to the effect of a refraction index.
To see how this arrives, we need to go back to the effective electroweak Lagrangian in Eq.~\eqref{eq:effective-weak-lagrangian}.
To get the effective potential induced by matter, we trace over the background in which neutrinos propagate, that is, we integrate $\mathcal{L}_{\rm weak}^{\rm eff}$ in the dense medium.
Usual matter only has protons, electrons and neutrons. 
Neglecting all other fermions and focusing on the terms which contains a neutrino current, the relevant terms of the effective Lagrangian are
\begin{align}
  \mathcal{L}_{\rm weak}^{\rm eff} \supset & -2\sqrt{2}G_F \Bigg\{(\bar\nu_e\gamma_\mu P_L \nu_e)(\bar e\gamma^\mu P_Le) \\
 & \hspace{2cm}+ \left(\sum_\alpha \bar\nu_\alpha\gamma_\mu P_L \nu_\alpha \right)\left[\sum_f^{p,n,e}\bar f\gamma^\mu (I_3^f P_L - \sin^2\theta_W Q_f) f\right]\Bigg\},\nonumber
\end{align}
where a Fierz identity was used in the first term.

Defining $f_{e,p,n}(p)$ to be the electron, proton and neutron momentum distributions in the medium normalized as 
\begin{equation}
   \int d^3p \,f_{e,p,n}(p)=1,
\end{equation} 
and assuming the medium to be isotropic (and thus only the terms proportional to $\gamma^0$ are nonzero),
allows us to trace over the background $\ket\Omega$
\begin{align}
  \left\langle\mathcal{L}_{\rm weak}^{\rm eff}\right\rangle = &\,-2\sqrt{2}G_F \Bigg\{(\bar \nu_e\gamma^\mu P_L\nu_e) \int d^3p f_e(p) \bra{\Omega} \bar e\gamma^\mu P_L e \ket{\Omega}  \nonumber\\
  &\hspace{1cm}+\frac12 (\bar \nu_\alpha\gamma^\mu P_L\nu_\alpha) \int d^3p \bra{\Omega} 
      \Big[f_e(p) \bar e\gamma^\mu \left(- P_L +2s_W^2\right) e \nonumber\\
  &   \hspace{3cm} + f_p(p) \bar p\gamma^\mu \left(P_L -2s_W^2\right) p   + f_n(p) \bar n\gamma^\mu \left(-P_L\right) n \Big]
      \ket{\Omega}\Bigg\}\nonumber\\
      = &-\sqrt2G_F n_e (\bar \nu_e\gamma^0 P_L\nu_e) 
      - \frac{G_F}{\sqrt{2}}\left[(1-4s^2_W)(n_p-n_e)-n_n\right] (\bar \nu_\alpha\gamma^0 P_L\nu_\alpha),
\end{align}
where a sum on the flavor index $\alpha$ in the last term is implicit.
The number densities $n_{e,p,n}$ are obtained by noting that $\bar e_L\gamma^0 e_L = e_L^\dagger e_L$ is the number operator for left-handed electrons, and similar for protons and neutrons.

We define the charged current effective potential as 
\begin{equation}
  V_{\rm CC} = \sqrt{2}G_F n_e,
\end{equation}
and the neutral current one as  
\begin{equation}
  V_{\rm NC}=\frac{G_F}{\sqrt{2}}\left[(1-4\sin^2\theta_W)(n_p-n_e)-n_n\right].
\end{equation}
In the flavor basis, we can write the matter potential matrix as
\begin{equation}
  V = \text{diag}(V_{\rm CC} + V_{\rm NC}, \,V_{\rm NC}, \,V_{\rm NC}).
\end{equation}
Nevertheless, since a universal and diagonal contribution to the Hamiltonian does not change oscillation physics, as we have seen above, we can drop the neutral current contribution from the matter potential to evaluate oscillations of standard neutrinos in a dense background.
This leads to the effective neutrino Dirac equation
\begin{equation}
  (i\psl - m - \gamma^0 V)\nu(x) = 0,
\end{equation}
where the generation index is implicit (and thus $m$ and $V$ are matrices). 
Here, it is easier to work in the mass basis.
Substituting the neutrino spinor by its plane-wave solution $\nu_i(x)=u_i(p)e^{-i p_i\cdot x}$ and using the fact that $(\slash\!\!\!p_i-m_i)u_i(p)=0$, we can arrive at
\begin{equation}
  \left[i\partial_t-(E_i \delta_{ij}+V^0_{ij})\right]\psi_j(x)=0.
\end{equation}
We can then write the neutrino evolution in the mass basis as
\begin{equation}\label{eq:neutrino-evolution-matter}
  i\frac{\partial}{\partial t}\psi_i = \left(\frac{\Delta m^2_{i1}}{2p}\delta_{ij}+V^0_{ij}\right)\psi_j,
\end{equation}
with $V^0 = U V U^\dagger$. 
For antineutrinos, since they correspond to negative energy solutions of the Dirac equation, there is a flip in the  relative sign of $V$ in Eq.~\eqref{eq:neutrino-evolution-matter}. 
This change in sign changes the oscillation probability of neutrinos with respect to antineutrinos and is key to determine the ordering of the neutrino mass eigenstates, as we will see shortly.

It is useful to analyze what happens in a simplified two neutrino scenario.
The Hamiltonian with matter effects would be written in the flavor basis as
\begin{equation}
  H = \frac{1}{2E}U^T\left(\begin{array}{cc}
 0 & 0 \\
 0  & \Delta m^2 
\end{array}
\right)U
+ \left(\begin{array}{cc}
  V & 0\\
  0&0
\end{array}
\right),
\end{equation}
where $U$ is a rotation in two dimensions
\begin{equation}
  U = \left(\begin{array}{cc}
  \cos\theta & \sin\theta\\
  -\sin\theta& \cos\theta
\end{array}
\right).
\end{equation}
It is straightforward to obtain the angle that diagonalizes this Hamiltonian $\tilde H = \tilde U H\tilde U^T$ and the new mass splitting,
\begin{align}
  \Delta \tilde m^2 &= \sqrt{A^2 - 2A\Delta m^2\cos(2\theta)+(\Delta m^2)^2},\\
  \tan (2\tilde\theta) &= \frac{\Delta m^2 \sin(2\theta)}{\Delta m^2 \cos(2\theta)-A},  \label{eq:mixing-matter}
\end{align}
where $A\equiv 2EV$.

\begin{figure}[t]
\centering
\includegraphics[width=0.95\linewidth]{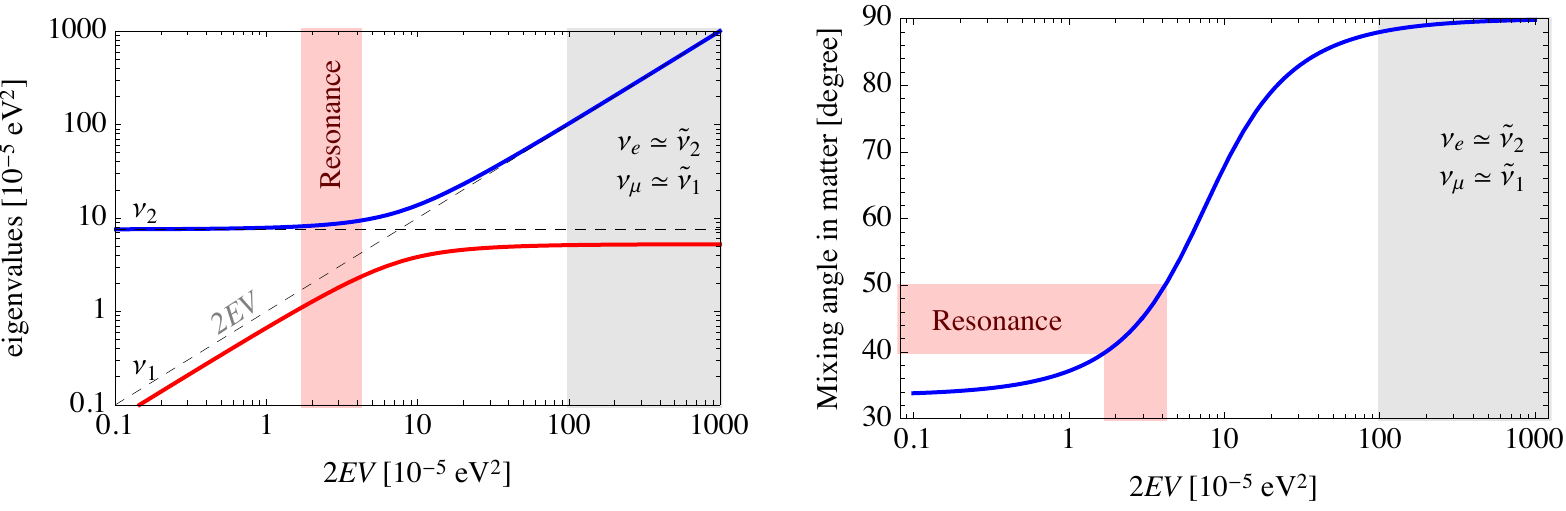}
\caption{Left: level crossing diagram for neutrinos in a matter background. Right: mixing angle as a function of the matter potential. For both panels we assumed a simplified two neutrino framework with $\Delta m^2_{21}=7.5\times10^{-3}$~eV$^2$ and $\sin^2\theta=0.3$ in vacuum.
\label{fig:MSW}}
\end{figure}

We can obtain some physics insight on the effect of matter on neutrino oscillations by looking at Fig.~\ref{fig:MSW}. In the left panel, we present the level crossing diagram, highlighting the resonance region, which will be defined soon, in red and the very dense regime in gray.
In the right panel, we show the value of the mixing angle in matter as a function of the matter potential, highlighting the same regions.
For this discussion we have used the measured values of the solar parameters, $\Delta m^2_{21} = 7.53\times 10^{-5}$~eV$^2$ and $\sin\theta_{12}=0.304$, see Sec.~\ref{sec:kamland-solar}.
The MSW resonance happens when the system reaches maximal mixing angle in matter, $\tilde\theta=45^\circ$.
Using Eq.~\eqref{eq:mixing-matter}, we can see this happens for 
\begin{equation}
  2EV_{\rm res} = \Delta m^2\cos(2\theta).
\end{equation}
Thus, we have arbitrarily defined the resonance region as that in which the mixing angle in matter is between $40^\circ$ and $50^\circ$.

If we label the flavor eigenstates by $\ket{\nu_e} = c_\theta \ket{\nu_1} + s_\theta \ket{\nu_2}$ and $\ket{\nu_\mu} = -s_\theta \ket{\nu_1} + c_\theta \ket{\nu_2}$, and similarly for the matter eigenstates $\tilde\nu_{1,2}$, we see that in a very dense medium, $A\gg \Delta m^2 \cos(2\theta)$, we have $\nu_e\simeq \tilde \nu_2$ and $\nu_\mu\simeq\tilde\nu_1$, as the angle is close to $90^\circ$.
We also see that in vacuum, $A\ll \Delta m^2 \cos(2\theta)$, we recover $\tilde U=U$ as expected.
If we focus on the Sun, which via nuclear fusion produces neutrinos around the MeV scale and whose core has electron number densities of order $100\, N_A/{\rm cm}^3$, where $N_A=6.02\times 10^{23}$ is Avogadro's number, we can write the resonance condition as
\begin{equation}
  \left(\frac{E}{\rm MeV}\right)\left(\frac{n_e}{100 N_A/{\rm cm}^3}\right)\simeq 1.9 \left(\frac{\Delta m^2}{7.53\times10^{-5}~{\rm eV}^2}\right)\left(\frac{\cos(2\theta)}{0.39}\right).
\end{equation}
While low energy solar neutrinos have energies below the MeV, and thus oscillate essentially in vacuum, higher energy solar neutrinos can reach 10~MeV or so, well above the MSW resonance condition.

\begin{figure}[t]
\centering
\includegraphics[width=0.95\linewidth]{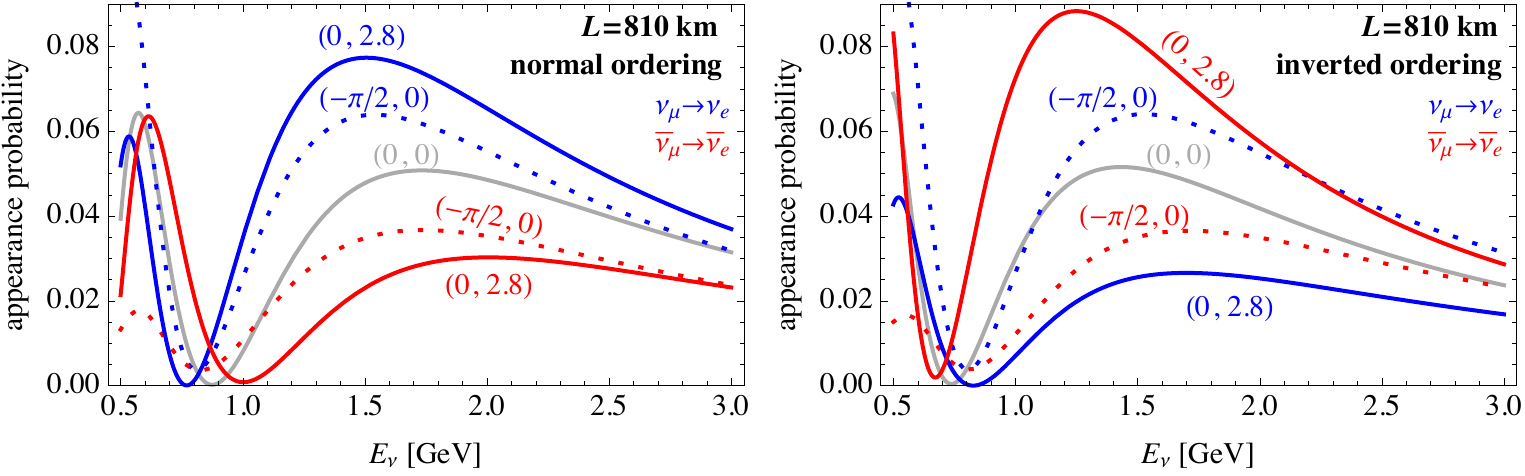}
\caption{Electron neutrino appearance probability for the normal (left panel) and inverted (right panel) mass orderings for neutrinos (blue) and antineutrinos (red) several values of $\delta_{CP}$ and matter density, as indicated in the labels as $(\delta_{CP},\,\rho/{\rm cm}^3)$. 
The $(0,\,0)$ lines for neutrinos and antineutrinos are identical and depicted in gray. 
We see that $\delta_{CP}$ has a similar effect on the appearance probabilities regardless of the ordering (see dotted lines), while the effect of the matter potential strongly depends on the neutrino mass spectrum (see solid lines).
\label{fig:oscillations-matter}}
\end{figure}

It is important to note that matter effects are sensitive to the sign of the mass splitting and that the MSW resonance can only happen for neutrinos if $\Delta m^2 \cos(2\theta)$ is positive, and for antineutrinos if $\Delta m^2 \cos(2\theta)$ is negative.
In other words, the effect of the matter potential depends on the ordering of the mass eigenstates.
Moreover, since the $CP$ violating phase also changes the oscillation probabilities of neutrinos with respect to antineutrinos, a careful planning of experimental baselines and energies may yield crucial information on the neutrino mass ordering and $CP$ violation.
To exemplify this point, we show in Fig.~\ref{fig:oscillations-matter} the appearance probability for $\nu_\mu\to\nu_e$ (blue lines) and $\bar\nu_\mu\to\bar\nu_e$ (red lines) for a baseline of 810~km, varying the values of $\delta_{CP}$ and the matter density.
The values used are indicated as $(\delta_{CP},\,\rho/{\rm cm}^3)$, assuming a 1:1:1 ratio of electrons, protons and neutrons.
The $(0,\,0)$ curve is identical for neutrinos and antineutrinos, and is shown in gray.

In the left panel we have assumed normal mass ordering, $\Delta m^2_{31}>0$, while the right panel was drawn assuming inverted ordering, $\Delta m^2_{31}<0$.
Comparing the dotted lines to the gray one, we see that the impact of $\delta_{CP}$ on the neutrino and antineutrino appearance is not much dependent on the mass ordering.
On the other hand, comparing the solid lines to the gray line, we can see that the presence of matter effects enhance the neutrino appearance and suppress antineutrino appearance in the normal ordering and does the opposite for inverted ordering.
Not mentioned here, but also relevant, the mixing angle $\theta_{23}$ also plays a crucial role in the appearance probabilities, but we will not discuss it in detail.
The take home message here is that the determination of the neutrino mass ordering via electron neutrino appearance also involves measuring, to some extent, the $CP$ phase and $\theta_{23}$.
The explicit, approximate formulae for oscillation probabilities for several different baselines and channels will be discussed in Sec.~\ref{sec:measurements} when we present the measurements of oscillation parameters.
This will allow us to better identify the dependence of relevant oscillation probabilities on the mass splittings, mixings and $CP$ phase.

As a last comment, for constant matter density, that is, $V(L)=V$, the oscillation probability for neutrinos propagating in matter can be calculated similarly to what was done in Eqs.~(\ref{eq:transition-amplitude}-\ref{eq:transition-amplitude-final}), but now adding the matter potential to the Hamiltonian and diagonalizing $H+V$ in order to exponentiate it.
In a two neutrino framework, the disappearance oscillation probability for constant matter density, for example,  would be given by
\begin{equation}
  P(\nu_\alpha\to\nu_\alpha) = 1 - \sin^2(2\tilde\theta)\sin^2\left(\frac{\Delta \tilde m^2 L}{4E}\right).
\end{equation}
For variable matter density, the evolution equation could be solved as any system of coupled differential equations.
In practice it can be convenient to discretize the changes in the matter potential and evolve the system ``slice per slice,'' that is
\begin{equation}
  P(\nu_\alpha\to\nu_\beta;L)=|\mathcal{A}_{\alpha\beta}|^2=\left|\bra{\nu_\beta}e^{-i H_N \Delta L}..e^{-i H_k \Delta L}..e^{-i H_1 \Delta L}\ket{\nu_\alpha}\right|^2,
\end{equation}
where $H_k$, $k=1..N$, is the full evolution operator for the $k$-th slice and $\Delta L\equiv L/N$.

\subsection{Density matrix formalism}

Some astrophysical environments, such as in supernovae, may have such dense media that the interaction of neutrinos with background particles cannot be neglected in the evolution of system.
To properly take those effects into account, it is convenient to introduce the neutrino evolution within the density matrix formalism.
We refer the reader to Ref.~\cite{gottfried2003quantum} for a more in-depth introduction to the density matrix formalism.
In general terms, the density matrix is defined as the weighted outer product of the state that defines a system.
For instance, if the eigenvalues of the Hamiltonian that describes the system are energy values $E$ and there is some other observable compatible with the Hamiltonian, say $k$, then the density matrix can be written as
\begin{equation}
  \rho = \sum_{Ek}\ket{Ek}f(E)\bra{Ek},
\end{equation}
where $f(E)$ describes a probability distribution.
A common example would be a system in thermal equilibrium in which the $f(E)$ is given by, e.g., the Boltzmann probability distribution $f(E)=N e^{-E/T}$, where $N$ is a normalization $N=\sum_E e^{-E/T}$.
As probabilities are normalized the density matrix obeys ${\rm Tr}(\rho) = 1$.
If the system is in a pure state, $\rho = \ket\psi\bra\psi$, we have ${\rm Tr}(\rho^2) = 1$, else ${\rm Tr}(\rho^2) < 1$.
Finally, the expectation value of an observable $Q$ is given by $\langle Q \rangle={\rm Tr}(\rho Q)$.

The evolution of the density matrix can be logically separated into two terms, one concerning the free part of the Hamiltonian, and another which accounts for interactions.
The free Hamiltonian is hermitian. 
As we have seen in Eq.~\eqref{eq:neutrino-evolution-matter}, the free Hamiltonian evolution of a (neutrino) state is simply $i\partial_t\ket{\psi}=H\ket{\psi}$, where  $H$ now denotes the Hamiltonian including the refraction index due to matter effects. 
The density matrix evolution in a collisionless environment is given by 
\begin{equation}
  ~~~~~~~~~~~~~~~~~~~~~~~~~~~~~~~~i \dot{\rho} = [H,\rho]\hspace{1cm}\text{(collisionless case).}
\end{equation}
To include the effect of collisions, we note that scatterings are modeled by the imaginary part of the optical potential, say $C$, and thus the evolution of $\ket{\psi}\bra{\psi}$ gets an additional sign, leading to
\begin{equation}
  i \dot{\rho} = [H, \rho] + \{C,\rho\}.
\end{equation}

In this formalism one can write down the evolution of neutrinos, properly taking into account oscillations and collisions with the background. 
This is highly relevant in the early universe cosmology and in extreme environments such as supernovae, where the density of particles is so high that neutrino interactions cannot be neglected.
The physics of supernovae and early universe will be discussed in other chapters of this book.

\section{Oscillation measurements}\label{sec:measurements}

The establishment of the oscillation phenomenon in the neutrino sector is one of the major accomplishments of the previous decades in high energy physics. 
Because weak interaction cross sections at low energies are suppressed by $G_F^2$, neutrino experiments require very intense sources, exceptionally large detectors, and long exposures to gather enough data that may allow for significant measurements of neutrino parameters.
In fact, the establishment of the neutrino sector was only possible due to a heroic effort performed by a number of experiments since the discovery of the neutrino by the Cowan--Reines experiment~\cite{Cowan:1992xc}, the discovery of a second type of neutrino by Lederman, Schwartz and Steinberger~\cite{Danby:1962nd}, and the resolution of the solar neutrino problem found by Davis~\cite{Davis:1964hf}.

A short treatment of the history of neutrino physics and in particular neutrino oscillations would not do justice to such a fascinating moment in the history of high energy physics.
In view of that, we will discuss the current experiments that provide the most important measurements of oscillation parameters and highlight some near future experiments that will further our understanding of neutrino physics.
As mentioned above, neutrino oscillation phenomenology in the standard model can be completely described by two mass splittings and the PMNS matrix, which can be parametrized by three mixing angles and a $CP$ violating phase (the Majorana phases do not affect oscillations).

\subsection{KamLAND and solar neutrinos}\label{sec:kamland-solar}
Let us start with the measurements of the so-called solar neutrino parameters $\Delta m^2_{21}$ and $\theta_{12}$.
The solar mass splitting is best determined by the KamLAND experiment~\cite{Araki:2004mb}.
The KamLAND detector consists of 1~kton of ultra-pure liquid scintillator detector capable of observing electron antineutrinos via inverse beta decay on free protons, $\bar\nu_e p^+\to e^+ n$.
The experiment can detect the delayed ($\sim200~\mu$s) 2.2~MeV photon produced by neutron capture in the detector, which is used to suppress backgrounds.
Electron antineutrinos are copiously produced in several nuclear reactors spread throughout Japan, with a weighted average distance from KamLAND detector of about 180~km. 
As the spectrum of reactor antineutrinos peaks around 3~MeV and is significant up to about $8$~MeV or so, KamLAND probes $L/E\sim50$~km/MeV.
The survival probability relevant for KamLAND can be approximated by~\cite{Gando:2013nba}
\begin{equation}
  P(\bar\nu_e\to\bar\nu_e)\simeq \cos^4\theta_{13}\left[1-\sin^2(2\theta_{12})\sin^2\left(\frac{\Delta m^2_{21}L}{4E}\right)\right] + \sin^4\theta_{13}.
\end{equation}
Matter effects are small and were neglected in the formula above.

\begin{figure}[t]
\centering
\includegraphics[width=0.7\linewidth]{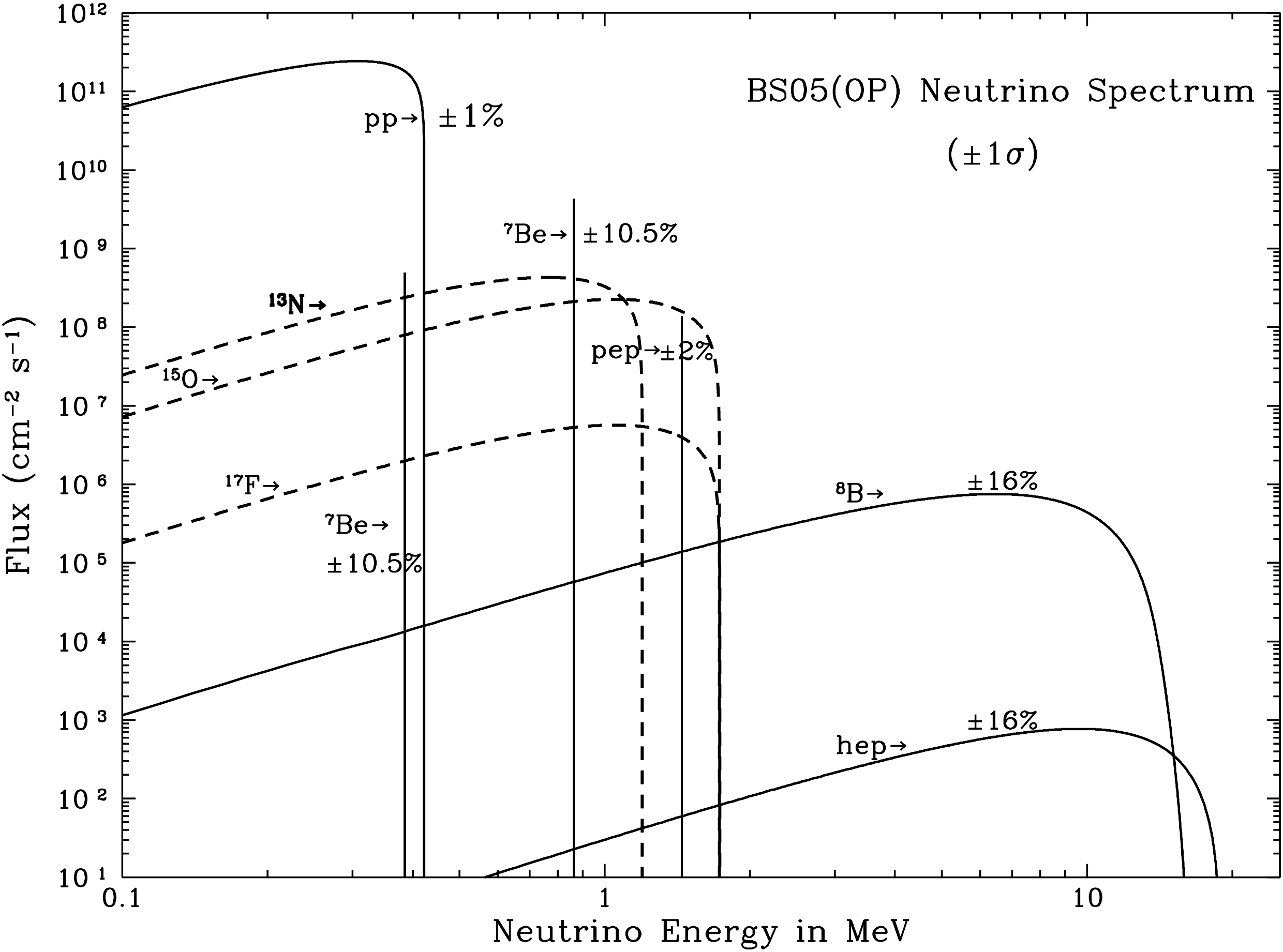}
\caption{Flux of solar neutrinos taken from Ref.~\cite{Bahcall:2004pz}.
\label{fig:solar-neutrinos}}
\end{figure}

The measurement of the solar mixing angle $\theta_{12}$, on the other hand, is dominated by solar neutrino experiments~\cite{Aharmim:2011vm,Abe:2016nxk,yasuhiro_nakajima_2020_3959640}. 
Solar neutrinos are mainly produced by two nuclear fusion chain reactions in the Sun: the $pp$ cycle which produces the vast majority of solar neutrinos and the CNO cycle which is connected to the metallicity of the Sun. 
The neutrino flux produced in the Sun can be found in Fig.~\ref{fig:solar-neutrinos}.

Some of the neutrinos produced in the Sun, particularly those from the decay of $^8$B, are strongly affected by matter effects, while lower energy neutrinos such as those in the $pp$ chain, essentially undergo oscillations in vacuum, see Sec.~\ref{sec:oscillations-in-matter}.
The role of matter effects depend on the sign of the mass splitting, and thus solar neutrinos are sensitive to the sign of $\Delta m^2_{21}$.
The full treatment of solar neutrino oscillations, accounting for adiabaticity~\cite{Parke:1986jy}, is beyond the scope of this chapter, but it is still useful to analyze two limiting scenarios: low energy solar neutrinos such as $pp$ neutrinos, $2EV\ll \Delta m^2_{21}\cos(2\theta_{12})$; and high energy solar neutrinos such as those from $^8$B decay, $\Delta m^2_{21}\cos(2\theta_{12}) \ll 2EV$.

For the first case, neutrinos oscillate freely in vacuum, and eventually decohere, leading to an oscillation probability of $P(\nu_e\to\nu_e) = \sum_i |U_{ei}|^4$.
In contrast, $^8$B neutrinos are produced well above the MSW resonance, and thus are approximately $\tilde\nu_2$, where the tilde represents the eigenstate in matter. 
As the evolution throughout the Sun is adiabatic, the system remains in the $\nu_2$ state as it reaches the vacuum. 
Thus, the disappearance probability would be approximately the $\nu_e$ admixture in $\nu_2$, that is $P(\nu_e\to\nu_e) \simeq |U_{e2}|^2$.
To be more precise and account for three neutrino oscillation effects, the relevant survival probabilities for such lower and higher energy solar neutrinos are given by
\begin{align}
  P(\nu_e\to\nu_e) &\simeq \sum_{i=1}^3 |U_{ei}|^4=c^4_{13}\left(c^4_{12}+s^4_{12}\right) + s^4_{13} &\text{(low energy solar $\nu$)},\\
  P(\nu_e\to\nu_e)&\simeq c^4_{13}s^2_{12} + s^4_{13}& \text{(high energy solar $\nu$)}.
\end{align}
Moreover, the matter potential due to neutrinos crossing the Earth at night, when compared to neutrinos observed during the day, displays a small but non-negligible effect on the oscillation probability.
This effect is referred to as the day-night asymmetry and it contributes to the determination of solar parameters.

Combining solar neutrino and KamLAND measurements, allows for the best determination of the solar parameters to date~\cite{Gando:2013nba},
\begin{equation}
  \Delta m^2_{21}=7.53^{+0.18}_{-0.18} \times 10^{-5}~{\rm eV}^2, 
  ~~~~~~~~~~\sin^2\theta_{12} = 0.304^{+0.014}_{-0.012},
\end{equation}
where the uncertainties are quoted at 68\% confidence level.
It is also worth highlighting that $\theta_{13}$ plays a small role in both KamLAND and solar neutrinos, and thus the inputs from short baseline reactor neutrino experiments such as Daya Bay and RENO, which will be discussed next, are quite relevant for a precise determination of solar parameters.
The determination of solar oscillation parameters will be further improved to the sub-percent level by the upcoming JUNO experiment~\cite{An:2015jdp}, a reactor neutrino detector with a 53~km baseline.
JUNO also proposes to measure the neutrino mass ordering by observing the frequency of the fast oscillations induced by the atmospheric mass splitting.

\subsection{Daya Bay and RENO}
Perhaps the most clean measurement of oscillation parameters is that of $\theta_{13}$ and the atmospheric mass splitting by the short baseline reactor neutrino experiments Daya Bay and RENO.
Those experiments are quite similar to the previously discussed KamLAND: liquid scintillator detectors probing electron antineutrinos from nuclear reactors.
Both experiments are also capable of detecting the delayed 2.2~MeV photon from neutron capture resulting from the inverse beta decay process to reduce backgrounds.
The main difference with respect to KamLAND is that the short baseline reactor experiments are located at about a kilometer or so from the nuclear reactor core.
That makes them sensitive to the atmospheric mass splitting and the smallest leptonic mixing angle, $\theta_{13}$.

To be more precise, the disappearance oscillation probability for that case can be approximated by~\cite{Nunokawa:2005nx}
\begin{equation}\label{eq:reactor-neutrino-prob}
  P(\bar\nu_e\to\bar\nu_e)\simeq 1-\sin^2(2\theta_{13})\sin^2\left(\frac{\Delta m^2_{ee} L}{4E}\right),
\end{equation}
where $\Delta m^2_{ee}\equiv c_{12}^2\Delta m^2_{31}+s_{12}^2\Delta m^2_{32}$.
Daya Bay provides the following measurements~\cite{Adey:2018zwh},
\begin{align}
  &\Delta m^2_{32} = 2.471^{+0.068}_{-0.070}\times 10^{-3}~{\rm eV}^2  & \text{(normal ordering)},\\
  &\Delta m^2_{32} = -2.575^{+0.068}_{-0.070}\times 10^{-3}~{\rm eV}^2 & \text{(inverted ordering)},\\
  &\sin^2(2\theta_{13}) = 0.0856\pm0.0029,
\end{align}
while RENO presents as~\cite{Bak:2018ydk}
\begin{align}
  &|\Delta m^2_{ee}| = \left[2.68\pm0.12({\rm stat})\pm0.07({\rm syst})\right]\times 10^{-3}~{\rm eV}^2, \\
  &\sin^2(2\theta_{13}) = 0.0896\pm0.0048({\rm stat})\pm0.0047({\rm syst}).
\end{align}
Alone, short baseline reactor experiments are not sensitive to the neutrino mass ordering, but when  combined with accelerator neutrinos, which will be discussed next, they to add non-trivial information on the atmospheric mass splitting.

\subsection{NOvA, T2K}
We proceed now to accelerator neutrino measurements.
The experiments that currently dominate the sensitivity are NOvA and T2K.
In accelerator neutrino experiments, neutrinos are produced by impinging protons on a target, which creates large amounts of pions, as well as other hadrons.
The charged mesons are focused by a magnetic horn system and propagate through a decay pipe.
The decays of charged mesons generate the neutrino flux.
Beam neutrinos consist mostly of muon neutrinos, due to the largest branching ratio of charged pion decays, with small but significant electron neutrino contamination.
Moreover, by selecting the polarity of the magnetic field, one can choose to focus positive or negative mesons.
This allows to have a neutrino beam dominated by either $\nu_\mu$ or $\bar\nu_\mu$.
The incoming neutrino energy threshold for muon production via CC interactions is approximately the muon mass, see Eq.~\eqref{eq:threshold}.
Therefore, the neutrino flux in accelerator neutrino experiments typically goes from a couple hundred MeV up to several GeV or more.
Because of that, the necessary baseline to observe oscillations is of hundreds of kilometers, e.g., 295~km for T2K and 810~km for NOvA.

Due to higher energy and magnetic focusing system, accelerator neutrino experiments can probe several oscillation channels.
The $\nu_\mu\to\nu_\mu$ and $\bar\nu_\mu\to\bar\nu_\mu$ disappearance channels are mainly sensitive to the atmospheric mass splitting and $\sin^2(2\theta_{23})$, with oscillation probability approximately given by
\begin{equation}\label{eq:numutonumu}
  P(\nu_\mu\to\nu_\mu)\simeq 1-\sin^2(2\theta_{23})\sin^2\left(\frac{\Delta m^2_{\mu\mu}L}{4E}\right),
\end{equation}
where $\Delta m^2_{\mu\mu}\equiv s_{12}^2\Delta m^2_{31}+c_{12}^2\Delta m^2_{32} + \sin(2\theta_{12})s_{13}\tan\theta_{23}\Delta m^2_{21}\cos\delta$ and matter effects were neglected. 
Note that the last term in $\Delta m^2_{\mu\mu}$ is parametrically smaller than the other two.
The combination of the measurements of the oscillation frequencies $\Delta m^2_{ee}$ by reactor neutrinos, see Eq.~\eqref{eq:reactor-neutrino-prob}, with  $\Delta m^2_{\mu\mu}$ by accelerator neutrinos and the solar splitting $\Delta m^2_{21}$ from solar experiments is one of the ways of determining the neutrino mass ordering~\cite{Nunokawa:2005nx}.

The appearance oscillation probabilities for $\nu_\mu\to\nu_e$ and $\bar\nu_\mu\to\bar\nu_e$ are a bit more  complicated, as matter effects play an important role there.
Nevertheless, it is still useful to at least look the most relevant terms of this oscillation probability in vacuum, namely 
\begin{equation}\label{eq:numutonue}
  P(\nu_\mu\to\nu_e) = \sin^2(2\theta_{13})s^2_{23}\sin^2\left(\frac{\Delta m^2_{31}L}{4E}\right) -8J\sin\Delta_{21}\sin\Delta_{31}\sin\Delta_{32} + \dots,
\end{equation}
where the Jarlskog invariant is $J = s_{23}c_{23}s_{13}c_{13}^2s_{12}c_{12}\sin\delta$ and we have defined $\Delta_{ij}\equiv \Delta m^2_{ij}L/4E$.
For antineutrinos, the term with the Jarlskog invariant flips sign.

We can see several interesting features in Eqs.~\eqref{eq:numutonumu} and \eqref{eq:numutonue}. 
While $\theta_{23}$ drives muon neutrino disappearance, there is a degeneracy for $\theta_{23}$ above or below $\pi/4$. 
While this may sound uninteresting for the non-expert, the octant of $\theta_{23}$ is directly related to the amount of muon versus tau flavor in the mass eigenstates. 
Therefore, determining the octant of $\theta_{23}$ could contribute to our understanding of the flavor structure of the standard model.
The appearance channel is proportional to $\sin^2\theta_{23}$ and thus can solve this degeneracy.
Note also that for $CP$ violation to occur, the Jarlskog invariant and the oscillation phases in the second term of Eq.~\eqref{eq:numutonue} to be nonzero. This requires all mixing angles to be different from zero and $\pi/2$, the $CP$ phase to be different from 0 and $\pi$, and all mass splittings to be nonzero.
As a last remark, what cannot be seen in the equations above is the effect of the matter potential.
For the appearance channels, this effect depends on the sign of $\Delta m^2_{31}$, making this channel also sensitive to the mass ordering.

The latest measurements of atmospheric parameters by NOvA, using the measurements of short baseline reactor neutrinos to constrain $\theta_{13}$ and KamLAND/Solar to constrain $\theta_{12}$, are given as~\cite{alex_himmel_2020_3959581}
\begin{align}
  &\Delta m^2_{32}=(2.41\pm0.07)\times10^{-3}~{\rm eV}^2 & \text{(NOvA, normal ordering)},\\
  &\sin^2\theta_{23} = 0.57^{+0.04}_{-0.03} &\text{(NOvA, normal ordering)},
\end{align}
with a $1\sigma$ preference for normal ordering and $1.2\sigma$ preference for upper $\theta_{23}$ octant.
The allowed region for the $CP$ phase is too large for any meaningful quote.
The equivalent for T2K is quoted as~\cite{Abe:2021gky}
\begin{align}
  &\Delta m^2_{32} = (2.45\pm0.07)\times 10^{-3}~{\rm eV}^2 & \text{(T2K, normal ordering)},\\
  &\sin^2\theta_{23}= 0.53^{+0.03}_{-0.04} & \text{(T2K, normal ordering)},
\end{align}
with a weak preference for normal ordering and upper octant of 89\% and 80\% posterior probability, respectively.
T2K data alone exclude $CP$ conservation at the $2\sigma$ level.
Although both T2K and NOvA prefer the normal mass ordering, due to a slight disagreement in the $CP$ phase their combination exhibits a preference for inverted mass ordering~\cite{Kelly:2020fkv,Esteban:2020cvm}.

\subsection{Super-Kamiokande and IceCube/DeepCore}
The last oscillation measurements we will discuss are those from Super-Kamiokande and IceCube/DeepCore atmospheric neutrino sample.
Atmospheric neutrinos are produced when cosmic rays hit Earth's atmosphere, producing mesons whose decay chain leads to neutrinos, predominantly $\nu_\mu$, $\nu_e$, $\bar\nu_\mu$, and $\bar\nu_e$.
Atmospheric neutrinos have a wide spectrum of energy, from below the 100~MeV scale up to multi-TeV.
As neutrinos propagate through the Earth, they undergo oscillations.
Matter effects play an important role here, particularly for neutrinos between 0.1-1~GeV and those around 6~GeV or so, where MSW resonances happen.
The oscillation probability of atmospheric neutrinos is more involved, as the matter density varies significantly when we compare the crust, mantle and core of the Earth.
Oscillation of atmospheric neutrinos above the GeV scale are sensitive to the atmospheric parameters $\Delta m^2_{31}$ and $\theta_{23}$, the mass ordering and $CP$ violation.

The Super-Kamiokande experiment, a 50~kton water Cherenkov detector in Japan, has collected atmospheric neutrino data for over 20 years. 
Its latest results provide~\cite{Takhistov:2020qhw}
\begin{align}
  &\Delta m^2_{32} = (2.40^{+0.11}_{-0.12})\times 10^{-3}~{\rm eV}^2  & \text{(Super-Kamiokande, normal ordering)},\\
  &\theta_{23} = 0.44^{+0.05}_{-0.02}  & \text{(Super-Kamiokande, normal ordering)},\\
  &\delta = 4.36^{+0.88}_{-1.39}  & \text{(Super-Kamiokande, normal ordering)},
\end{align}
with a less than $2\sigma$ preference for the normal mass ordering.
The other experiment that also provides invaluable information with atmospheric neutrinos is IceCube/DeepCore, an array of photomultipliers deployed deep in the ice of Antartica.
Although IceCube/DeepCore is only sensitive to higher energy neutrinos, its larger detector mass allows it to gather very large statistics.
The result of eight years of data taking is given as~\cite{Aartsen:2017nmd}
\begin{align}
  &\Delta m^2_{31}= (2.31^{+0.11}_{-0.13})\times 10^{-3}~{\rm eV}^2&\text{(IceCube/DeepCore normal ordering)},\\
  &\sin^2\theta_{23}=0.51^{+0.07}_{0.09}&\text{(IceCube/DeepCore normal ordering)},
\end{align}
with almost no sensitivity to the mass ordering.

\section{Open questions in neutrino physics}\label{sec:questions}

To date, several open questions and unsolved puzzles related to neutrino physics remain.
Due to its low interaction rate, the neutrino sector remains  the least well understood sector of the standard model.
In what follows, we will describe some of these issues and the discuss present efforts to address them.

\subsection{Absolute values of neutrino masses}
As mentioned previously, although in the standard model neutrinos are strictly massless, the observation of neutrino oscillations require a nonzero mass for at least two neutrinos due to the presence of two nonzero mass splittings. 
In fact, the earliest proposal to measure the mass of neutrinos was put forward by Enrico Fermi in his seminal paper on the four fermion operator which was the basis of weak interactions~\cite{Fermi:1934sk}.
The proposal was to perform a kinematical measurement of neutrino masses using $\beta$-decays. In a nutshell, the presence of nonzero neutrino masses change the endpoint of $\beta$-decay spectra. 
The most sensitive experiment performing such a measurement is the ongoing KATRIN experiment in Germany.
By measuring the electron spectrum of $\beta$-decay using an intense tritium source, KATRIN is able to put the most stringent limit on the effective electron neutrino mass~\cite{Aker:2019uuj}
\begin{equation}
  m_{\nu_e}^{\rm eff}\equiv \sum_i |U_{ei}|^2m_i < 1.1~{\rm eV} \hspace{1cm} (90\%{\rm C.L.}).
\end{equation}

Another promising way to measure neutrino masses is by cosmological observations.
In fact, cosmology constrains both the number of relativistic neutrinos in the early universe, as well as the sum of their masses.
We will defer this discussion to another chapter where the role of neutrinos in cosmology will be discussed in more detail.

\subsection{Neutrino mass mechanism}\label{sec:neutrino-mass-mechanism}
As discussed in Sec.~\ref{sec:standard-model}, the implementation of the Higgs mechanism as an explanation for neutrino masses requires the presence of a standard model singlet. 
It would be possible then to write a Majorana mass term of such a singlet, $M_R\bar\nu_R^c \nu_R$, which in turn would completely changes the mass generation for neutrinos and allow for the possibility that the light neutrinos are Majorana particles.
In fact, there is a large number of models that could provide neutrino masses, probably due to the lack of guidance from experimental observations.
The reason behind the smallness of neutrino masses vary from model to model:
they could be related to loop suppressions, extra dimensions, small lepton number violation, a suppression from a large scale, etc. 

\begin{figure}[t]
\centering
 \parbox{1.62in}{	\centering
	\begin{fmffile}{seesaw-1}
	\begin{fmfgraph*}(100, 65)
		\fmfleft{i1,i2}
		\fmfright{o1,o2}
		\fmflabel{$L$}{i1}
		\fmflabel{$H$}{i2}
		\fmflabel{$L^c$}{o1}
		\fmflabel{$H$}{o2}
		\fmf{fermion}{i1,v1}
		\fmf{plain}{v2,v3,v1}
		\fmfv{decoration.shape=cross, label=$\nu_R$}{v3}
		\fmf{fermion}{o1,v2}
		\fmf{dashes}{i2,v1}
		\fmf{dashes}{o2,v2}
	\end{fmfgraph*}
	\end{fmffile}\vspace{0.5cm}
 \figsubcap{1} \label{fig:seesaw-1}}
 \parbox{1.62in}{	\centering
	\begin{fmffile}{seesaw-2}
	\begin{fmfgraph*}(100, 65)
		\fmfleft{i1,i2}
		\fmfright{o1,o2}
		\fmflabel{$L$}{i1}
		\fmflabel{$H$}{i2}
		\fmflabel{$L^c$}{o1}
		\fmflabel{$H$}{o2}
		\fmf{fermion}{i1,v1}
		\fmf{fermion}{o1,v1}
		\fmf{dashes}{i2,v2,o2}
		\fmf{dashes, label=$\Delta$}{v1,v2}
	\end{fmfgraph*}
	\end{fmffile}\vspace{0.5cm}
 \figsubcap{2} \label{fig:seesaw-2}}
 \parbox{1.62in}{	\centering
	\begin{fmffile}{seesaw-3}
	\begin{fmfgraph*}(100, 65)
		\fmfleft{i1,i2}
		\fmfright{o1,o2}
		\fmflabel{$L$}{i1}
		\fmflabel{$H$}{i2}
		\fmflabel{$L^c$}{o1}
		\fmflabel{$H$}{o2}
		\fmf{fermion}{i1,v1}
		\fmf{plain}{v2,v3,v1}
		\fmfv{decoration.shape=cross, label=$\psi$}{v3}
		\fmf{fermion}{o1,v2}
		\fmf{dashes}{i2,v1}
		\fmf{dashes}{o2,v2}
	\end{fmfgraph*}
	\end{fmffile}\vspace{0.5cm}
 \figsubcap{3} \label{fig:seesaw-3}}
 \caption{\label{fig:seesaw}Feynman diagrams for seesaw type I, II and III, labeled as (a), (b), (c), respectively.}
\end{figure}
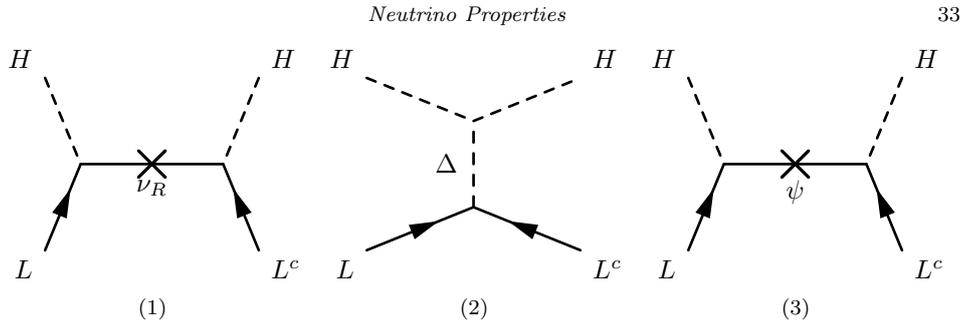

Perhaps the simplest way and most comprehensive way to generate neutrino masses is to use an effective field theory approach.
The lowest dimension operator that give rise to neutrino masses is the lepton number violating operator proposed by Weinberg~\cite{Weinberg:1979sa}
\begin{equation}\label{eq:weinberg-operator}
  \frac c\Lambda \left(\overline{L^c} \tilde H^*\right)\left(\tilde H^\dagger L\right),
\end{equation}
where $\Lambda$ is a scale and could be the one of neutrino mass generation, and $c$ is a Wilson coefficient which may has flavor indices.
It is curious that this is the only gauge-invariant dimension-5 operator that can be written in the standard model.
An entire class of models in which neutrinos are Majorana particles can be parametrized by Eq.~\eqref{eq:weinberg-operator}.
There are three simple models of neutrino masses that map into this operator, namely integrating out the following fields: a standard model singlet, $\nu_R$; a $SU(2)_L$ triplet scalar  with hypercharge $+1$, $\Delta=(\delta^0,\,\delta^+,\,\delta^{++})$; or a $SU(2)_L$ triplet fermion with hypercharge $0$, $\psi=(\psi^+,\,\psi^0,\,\psi^-)$. These are commonly known as seesaw type I, II and III, respectively~\cite{GellMann:1980vs,Yanagida:1979as,Mohapatra:1979ia,Schechter:1980gr,Magg:1980ut,Mohapatra:1980yp,Lazarides:1980nt,Ma:1998dx,Foot:1988aq,Ma:1998dn,Mohapatra:1986bd}, and are depicted diagrammatically in Fig.~\ref{fig:seesaw}.
A common feature of all these seesaw models is that neutrinos are Majorana particles, which bring us to another open problem of the standard model.

\subsection{Dirac vs. Majorana nature of neutrinos}

Since neutrinos are the only neutral, elementary fermions of the standard model, they are the only particles which could be Majorana.
A Majorana field is such that $\psi = e^{i\theta}\psi^c$, where $\theta$ is an unphysical phase, and was first proposed by Majorana~\cite{Majorana:1937vz}.
To better understand a Majorana field, it is useful to compare it to the usual Dirac fields.
A Dirac fermion is composed by four spinors which describe left- and right-handed particles and anti-particles.
If neutrinos were Dirac, weak gauge bosons  would interact with left-handed neutrinos and right-handed antineutrinos, while the other two degrees of freedom would be completely disconnected from the rest of the particles.

In contrast, a Majorana field has only two independent spinors, one of which is left-handed and another right-handed.
Due to the interactions with the $W$ boson, these would be identified with what we commonly refer to as neutrinos and antineutrinos, respectively.
It is not possible to define particles and antiparticles for a Majorana field, and it should be noted that referring to neutrinos and antineutrinos in this case is a misnomer.
For ``neutrinos'' to behave as ``antineutrinos,'' it would  require a chirality flip, which is typically suppressed by $m/E$ and thus negligibly small in almost all contexts.

\begin{figure}[t]
 \parbox{1.3in}{	\centering
	\begin{fmffile}{0nubb}
	\begin{fmfgraph*}(100, 65)
		\fmfleft{i1,i2}
		\fmfright{o1,o2,o3,o4}
		\fmflabel{$n$}{i1}
		\fmflabel{$n$}{i2}
		\fmflabel{$p^+$}{o1}
		\fmflabel{$e^-$}{o2}
		\fmflabel{$e^-$}{o3}
		\fmflabel{$p^+$}{o4}
		\fmf{fermion}{i1,v1}
		\fmf{plain}{v1,w1,o1}
		\fmf{phantom_arrow,tension=0}{v1,o1}
		\fmf{fermion}{i2,v5}
		\fmf{plain}{v5,w5,o4}
		\fmf{phantom_arrow,tension=0}{v5,o4}
		\fmf{plain}{v2,v3,v4}
		\fmfv{decoration.shape=cross, label=$\nu$}{v3}
		\fmf{plain}{v2,w2,o2}
		\fmf{phantom_arrow,tension=0}{v2,o2}
		\fmf{plain}{v4,w4,o3}
		\fmf{phantom_arrow,tension=0}{v4,o3}
		\fmf{photon, label=$W$}{v1,v2}
		\fmf{photon, label=$W$, label.side=left}{v4,v5}
	\end{fmfgraph*}
	\end{fmffile}\vspace{0.5cm}
 }\hspace{1.7cm}
 \parbox{1.8in}{	\centering
	\includegraphics[width=0.6\textwidth]{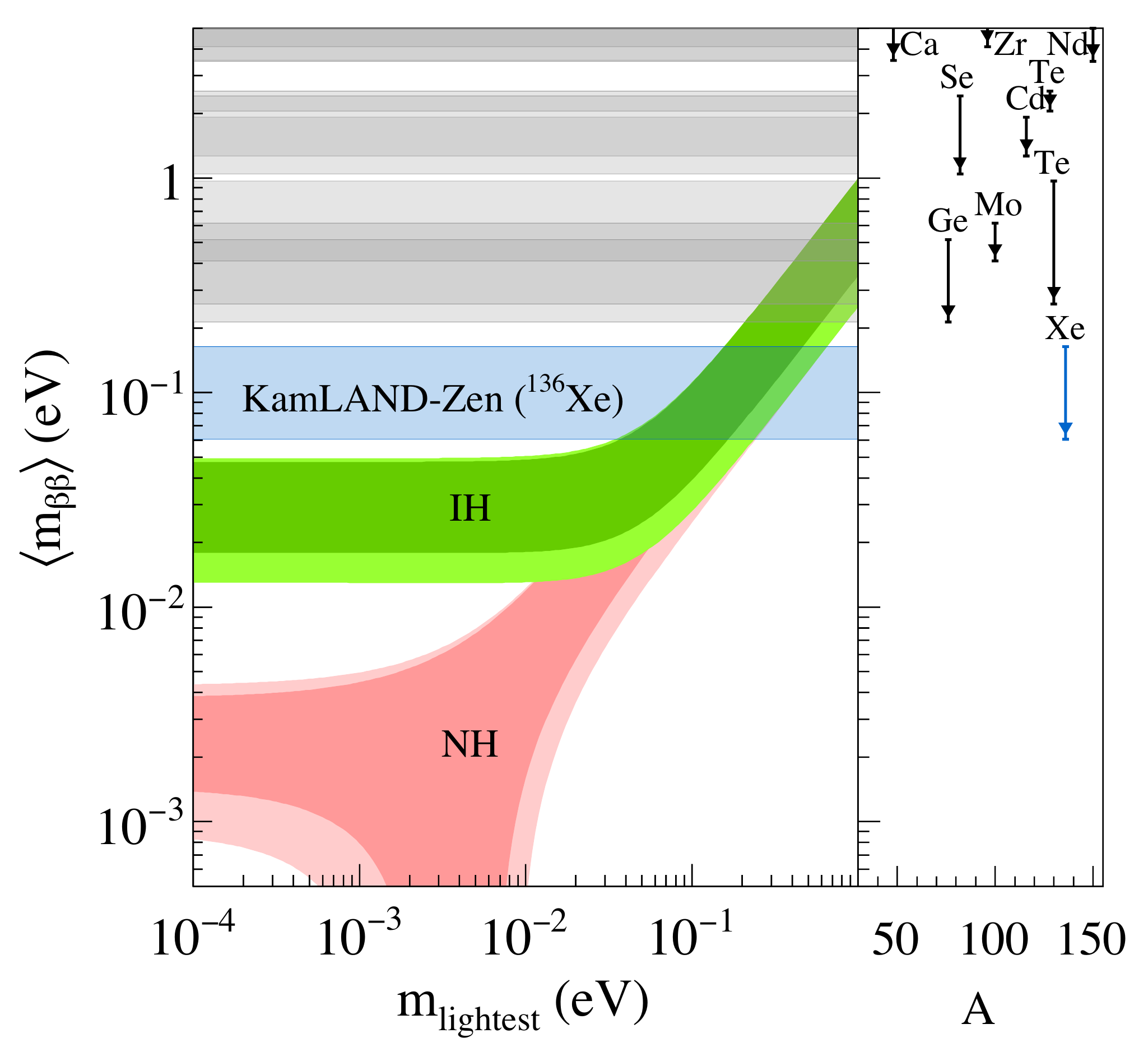}
  }
\caption{\label{fig:0nuBB}Left: Feynman diagram for neutrinoless double beta decay. Right: Allowed region for the effective mass $m_{\beta\beta} = |\sum_i U_{ei}^2m_i|$ given current knowledge of oscillation parameters for normal (red) and inverted (green) mass ordering. The darker regions are obtained by fixing oscillation parameters and varying only the Majorana phases, while the lighter regions allow for $3\sigma$ uncertainty variation of oscillation parameters. The KamLAND-Zen constraint in indicated as a blue band which encodes the uncertainty on the nuclear matrix element. the right panel shows limits from other experiments, highlighting the element used therein (see Ref.~\cite{KamLAND-Zen:2016pfg} for details).}
\end{figure}

The most striking feature of the possible Majorana nature of neutrinos is the violation of lepton number, and the most distinct observable that can probe it is the neutrinoless double beta decay ($0\nu\beta\beta$).
A nucleus with mass and atomic number $(A,Z)$ can undergo double $\beta$ decay when it has a heavier $Z+1$ isobar but a lighter $Z+2$ isobar.
In such case, the nucleus  cannot $\beta$-decay, but can ``double-$\beta$,'' that is, 
\begin{equation}
  (A,Z) \to (A,Z+2) + e^- + e^- + \bar\nu_e + \bar\nu_e,
\end{equation}
emitting two electrons and two antineutrinos.
For instance, this is the case for $^{76}_{32}$Ge which cannot beta decay to $^{76}_{33}$As but can double-beta decay to $^{76}_{34}$Se, and for the isobars $^{136}_{54}$Xe, $^{136}_{55}$Cs and $^{136}_{56}$Ba.
This process has been observed copiously and, apart from some nuclear physics aspects, it is well understood.
Nevertheless, if neutrinos are Majorana particles, the neutrino line can be virtual leading to a double $\beta$ decay without neutrinos, see left panel of Fig.~\ref{fig:0nuBB}.

Currently, the best constraint on the neutrinoless double beta decay lifetime of $^{136}$Xe comes from the KamLAND-Zen experiment in Japan which uses the refurbished KamLAND detector to look for these rare decays~\cite{KamLAND-Zen:2016pfg}.
The experiment has put the constraint on the $0\nu\beta\beta$ half-life of
\begin{equation}
  T^{0\nu}_{1/2} > 1.07 \times 10^{26}~{\rm years}\hspace{1cm}{\rm (90\%~C.L.)}.
\end{equation}
This can be translated into an effective double $\beta$ mass
\begin{equation}
  m_{\beta\beta} = \left| \sum_i U_{ei}^2m_i\right| < 61-165~{\rm meV},
\end{equation}
where the range correspond to nuclear matrix element uncertainties.
Note that the effective mass is sensitive to the Majorana phases of the PMNS matrix, as well as the $CP$ phase $\delta$.
Given the present knowledge of oscillation parameters, the allowed region and the KamLAND constraint on the effective mass $m_{\beta\beta}$ is presented in Fig.~\ref{fig:0nuBB}.
As a final comment, there are several variations of neutrinoless double beta decay that could be induced by violating lepton number by two units: double positron decay $(A,Z)\to(A,Z-2)+2e^+$; electron capture $e^- + (A,Z)\to(A,Z-2)+e^+$; and double electron capture $2e^- + (A,Z)\to(A,Z-2)$. 
These offer further opportunities to look for the Majorana nature of neutrinos.

\subsection{Short baseline experimental anomalies}

So far all open question discussed were rooted in theoretical considerations.
Nevertheless, there are experimental observations, apparently inconsistent with the three neutrino framework, that remain unexplained to date.
These are the anomalous excess of events observed by two experiments.
First, the Liquid Scintillator Neutrino Detector (LSND) at the Los Alamos National Laboratory  was deployed about 30~meters from an intense pion decay-at-rest source.
At LSND, pions were produced by impinging 800~MeV protons on a target, mostly $\pi^+$.
These in turn would stop at a beam dump and decay as $\pi^+\to\mu^+\nu_\mu\to\ e^+\nu_e\bar\nu_\mu\nu_\mu$.
The energy spectrum of those neutrinos is quite well known, averaging around 30~MeV.
The LSND experiment searched for $\bar\nu_e$ via the inverse beta process, by combining the Cherenkov, scintillation light yields, as well as the delayed light from neutron capture.
A significant excess of $87.9\pm22.4_{\rm stat}\pm6.0_{\rm syst}$ events was observed over the expected background~\cite{Aguilar:2001ty}, consisting of a $3.8\sigma$ anomaly.
If interpreted as neutrino oscillations, this short baseline $\bar\nu_\mu\to\bar\nu_e$ appearance would require a mass splitting of about 1~eV$^2$ or larger, see Eq.~\eqref{eq:two-flavor-oscillation}, and a mixing $\sin^2(2\theta)\sim 0.003$.
Since this cannot be explained by standard oscillations, it would require a fourth neutrino without weak interactions, a \emph{sterile neutrino}.

To test the sterile neutrino interpretation of the LSND anomaly, the MiniBooNE experiment was proposed.
It consisted of a neutrino beam line, the Booster Neutrino Beam, where 8~GeV protons hit a target, producing pions that would be focused by a magnetic horn system, which in turn would decay in a decay pipe.
The neutrino energy spectrum of MiniBooNE had energies around 1~GeV and consisted mostly of $\nu_\mu$ or $\bar\nu_\mu$ in the neutrino and antineutrino run modes, respectively.
These neutrinos would be detected at a mineral oil detector, located about 450 meters downstream the beam.
The MiniBooNE detector could distinguish muons from electrons, produced by charged current interactions of neutrinos, via the Cherenkov light pattern.
Recent results of the MiniBooNE collaboration exhibit a combined $\nu_e+\bar\nu_e$ excess of $638\pm132.8$ events over backgrounds, totalizing a $4.8\sigma$ discrepancy against expectations~\cite{Aguilar-Arevalo:2020nvw}.
The interpretation of these results in terms of sterile neutrinos is compatible with LSND.

Nevertheless, a problem with the sterile neutrino interpretation of the excesses arrives when considering other oscillation experiments.
In the sterile neutrino scenario, if the new mass splitting is much larger than the standard ones, the $\nu_\mu\to\nu_e$ appearance probability is connected to the $\nu_e$ and $\nu_\mu$ disappearance channels as
\begin{align}
  &P(\nu_\mu\to\nu_e)\simeq 4|U_{e4}U_{\mu4}|^2\sin^2\left(\frac{\Delta m_{41}^2 L}{4E}\right),\\
  &P(\nu_e\to\nu_e)\simeq 1- 4|U_{e4}|^2(1-|U_{e4}|^2)\sin^2\left(\frac{\Delta m_{41}^2 L}{4E}\right),\\
  &P(\nu_\mu\to\nu_\mu)\simeq 1- 4|U_{\mu4}|^2(1-|U_{\mu4}|^2)\sin^2\left(\frac{\Delta m_{41}^2 L}{4E}\right),
\end{align}
where the PMNS matrix has been expanded to four mass eigenstates.
Therefore, short baseline $\nu_\mu\to\nu_e$ appearance requires both $\nu_e$ and $\nu_\mu$ disappearance to be nonzero.
Experiments sensitive to the disappearance channels at short baselines, particularly MINOS/MINOS+~\cite{Adamson:2017uda} and IceCube~\cite{Aartsen:2020fwb} which measure $\nu_\mu$ disappearance, present a strong $4.7\sigma$ tension with the sterile neutrino interpretation of LSND and MiniBooNE anomalies~\cite{Dentler:2018sju}, see also Refs.~\cite{Gariazzo:2017fdh,Diaz:2019fwt}.
Besides LSND and MiniBooNE, there are two other anomalies in the neutrino sector, the reactor antineutrino anomaly~\cite{Mention:2011rk,Huber:2011wv} and the gallium anomaly~\cite{Giunti:2010zu}. These two anomaly stem from complicated theoretical calculations which are not completely understood. We will not address those here.

Besides all that, a sterile neutrino species with this relatively large mixing would be in thermal equilibrium during big bang nucleosynthesis.
This would lead to a larger number of effective relativistic degrees of freedom, which under the standard cosmology framework would be ruled out.
In a nutshell, to date, the excesses found by LSND and MiniBooNE are not understood.
There is no clear satisfactory standard model interpretation of the anomaly, such as unknown nuclear physics effects, neither beyond standard model interpretations that explain all experimental data without creating tension with other experiments and/or cosmological observations.

The Short Baseline Neutrino program at Fermilab, a set of three liquid argon time projection chamber (LArTPC) detectors deployed along the Booster Neutrino Beamline, is expected to clarify the nature of the MiniBooNE anomaly~\cite{Machado:2019oxb}. 
MiniBooNE cannot distinguish electron events from those originated from photon conversion, as the Cherenkov signal would be identical. LArTPCs on the other hand, should be able to distinguish electrons from photons due to the different ionization yield signature and gap between interaction vertex and the beginning of the electromagnetic shower.
This will clarify if the MiniBooNE excess is indeed caused by electrons or if photons are responsible for it.
On top of that, the combination of the three SBN detectors, namely, SBND, MicroBooNE and ICARUS, will be highly sensitive to electron neutrino appearance and muon neutrino disappearance in short baselines.
Together, they will be able to confirm or rule out the sterile neutrino interpretation of the LSND/MiniBooNE anomalies within a single experiment. 

\bibliographystyle{ws-rv-van}
\bibliography{bibliography}

\end{document}